\newif{\ifjournal} 
  \journaltrue 
\ifjournal 
    \documentclass[useAMS,usenatbib,usegraphicx]{mn2e} 
    \usepackage{graphicx,mathptm}

\else 
  \documentclass{aastex} 
\fi 
\input{psfig.sty}  
\usepackage{times} 
 
\def\lsim{\mathrel{\rlap{\lower4pt\hbox{\hskip1pt$\sim$}} 
    \raise1pt\hbox{$<$}}}                
\def\gsim{\mathrel{\rlap{\lower4pt\hbox{\hskip1pt$\sim$}} 
    \raise1pt\hbox{$>$}}}                

\def\2A     {{2A 0335+096}} 
\def\degd  {$^{\circ}\!$} 
 
\newcommand{\be}{\begin{equation}} 
\newcommand{\ee}{\end{equation}} 
\newcommand{\bed}{\begin{displaymath}} 
\newcommand{\eed}{\end{displaymath}} 
\newcommand{\ba}{\begin{eqnarray}} 
\newcommand{\ea}{\end{eqnarray}} 
  
\newcommand{\chandra}{\emph{Chandra}} 
\newcommand{\xmm}{XMM-\emph{Newton}} 
\newcommand{\tspec}{$T_{\rm spec}$} 
\newcommand{\tew}{$T_{\rm ew}$} 
\newcommand{\tsl}{$T_{\rm sl}$}

\begin{document} 
 
\title[Comparing  temperatures from hydro-N-body simulations to 
{\it Chandra} and XMM-{\it Newton} observations] 
{
Comparing the temperatures of galaxy clusters from hydro-N-body
simulations to {\it Chandra} and XMM-{\it Newton} observations
}
 
\author[P. Mazzotta et al.] 
{P. Mazzotta$^{1,2}$\thanks{E-mail: mazzotta@roma2.infn.it}, 
  E. Rasia$^3$, L. Moscardini$^4$, G. Tormen$^3$  
\\  
$^1$Dipartimento di Fisica, Universit\`a di
Roma "Tor Vergata", via della Ricerca Scientifica 1, I-00133 Roma, Italy\\ 
$^2$Harvard-Smithsonian Center for Astrophysics, 60 Garden 
Street, Cambridge, MA 02138, USA\\  
$^3$Dipartimento di Astronomia, Universit\`a di Padova, vicolo 
dell'Osservatorio 2, I-35122 Padova, Italy \\  
$^4$Dipartimento di Astronomia, Universit\`a di Bologna, via Ranzani 1,  
I-40127 Bologna, Italy  
} 
\maketitle 
 
\begin{abstract} 
Theoretical studies of the physical processes guiding the formation
and evolution of galaxies and galaxy clusters in the X-ray are mainly
based on the results of numerical hydrodynamical N-body simulations,
which in turn are often directly compared to X-ray observations.
Although trivial in principle, these comparisons are not always
simple. We demonstrate that the projected spectroscopic temperature of
thermally complex clusters obtained from X-ray observations is always
lower than the emission-weighed temperature, which is widely used in
the analysis of numerical simulations.  We show that this temperature
bias is mainly related to the fact that the emission-weighted
temperature does not reflect the actual spectral properties of the
observed source. This has important implications for the study of
thermal structures in clusters, especially when strong temperature
gradients, like shock fronts, are present.  Because of this bias, in
real observations shock fronts appear much weaker than what is
predicted by emission-weighted temperature maps, and may even not be
detected. This may explain why, although numerical simulations predict
that shock fronts are a quite common feature in clusters of galaxies,
to date there are very few observations of objects in which they are
clearly seen.  To fix this problem we propose a new formula, the
spectroscopic-like temperature function, and show that, for
temperature larger than 3 keV, it approximates the spectroscopic
temperature better than few per cent, making simulations more directly
comparable to observations.
\end{abstract} 
 
 
\section{Introduction} 
 
Clusters of galaxies have great cosmological relevance due to the
privileged role they play in the hierarchical scenario of cosmic
structure formation.  In fact they represent the largest cosmic
structures that had time to undergo gravitational collapse and
virialize.  This characteristics makes galaxy clusters a powerful tool
and a fundamental ingredient in many tests for the determination of
the main cosmological parameters, like the matter density $\Omega_m$
and the normalization and shape of the power spectrum of primordial
fluctuations (represented by the spectrum amplitude $\sigma_8$ and
spectral index $n$, respectively).
 
From a theoretical point of view, the relevant quantity to measure is
the cluster virial mass; this - however - can be directly determined
only through gravitational lensing measurements, and even so many
systematic effects make such measurement not an easy one.
Alternatively, virial masses can be estimated indirectly from related
quantities, as done with X-ray or millimetric (e.g. the
Sunyaev-Zel'dovich effect; \citealt{1972CoASP...4..173S}) observations,
which - under appropriate assumptions - link the cluster mass to its
temperature. Theoretical arguments suggest in fact the existence of
the so called scaling laws, tight correlations between mass and other
global quantities, in this case the mean cluster temperature
\citep{1986MNRAS.222..323K}.  Such correlations, confirmed and
calibrated by the results of hydrodynamical simulations
(see, e.g., \citealt{1995MNRAS.275..720N}; 
\citealt{1996ApJ...469..494E}; \citealt{1998ApJ...495...80B}; 
\citealt{1998ApJ...503..569E}), allow for
instance to link the observed cluster temperature function to the
theoretically estimated mass function.  Another, more detailed, method
to infer cluster masses is based on the solution of the equation of
hydrostatic equilibrium, where however an assumption on the spatial
temperature distribution is required (see, e.g.,
\citealt{1996ApJ...469..494E}; \citealt{1997MNRAS.286..865T};
\citealt{1998MNRAS.296.1061T}; \citealt{2003astro.ph..9405R}).
 
These arguments illustrate how important it is to have a reliable
description of the thermal structure of galaxy clusters.  This
description needs also be spatially detailed, as recent observational
data with high spatial and spectral resolution suggest that clusters
are far from isothermal, and show instead a number of peculiar thermal
features, like cold fronts (\citealt{2000ApJ...541..542M};
\citealt{2001ApJ...551..160V};
\citealt{2001ApJ...555..205M}; \citealt{2001ApJ...562L.153M};
\citealt{2003ApJ...596..190M}), 
cavities (\citealt{2000ApJ...534L.135M}; 
\citealt{2000MNRAS.318L..65F}; 
\citealt{2001ApJ...558L..15B}; \citealt{2001ApJ...562L.149M};
\citealt{2002ApJ...567L..37M}; 
\citealt{2001ApJ...562L.153M}; 
\citealt{2002MNRAS.336..299J}; 
\citealt{2002ApJ...569L..79H}; 
\citealt{2002ApJ...579..560Y};  
\citealt{2002MNRAS.331..273S}; 
\citealt{2002ApJ...565..195S}),  
blobs and filaments (\citealt{2001MNRAS.321L..33F};
\citealt{2002ApJ...569L..31M}; \citealt{2003ApJ...596..190M}).

A theoretical interpretation of
these observations clearly requires state-of-the-art hydro-N-body
simulations, which can be used to extract realistic temperature maps
and/or profiles.  The comparison between real and simulated data,
however, is complicated by different problems, produced both by
projection effects and by instrumental artifacts, like instrument
response, sky background and instrumental noise.  Finally, a further
complication can arise from a possible mismatch between the
spectroscopic temperature $T_{\rm spec}$ estimated from X-ray
observations and the temperatures usually defined in numerical
results.  In fact, while $T_{\rm spec}$ is a mean projected
temperature obtained by fitting a single or multi-temperature thermal
model to the observed photon spectrum, theoretical models fully
exploit the three-dimensional thermal information carried by gas
particles and so usually define physical temperatures.
 
For the above reasons, possible biases can naturally arise when
comparing the thermal structure of observed and simulated galaxy
clusters.
Using a set of 24 hydrodynamic-simulated clusters to obtain simulated
images with quality similar to that expected in real \chandra ~
observations, \cite{2001ApJ...546..100M} found that the spectroscopic
temperature is lower than the mass-weighted temperature by roughly 10
to 20 per cent (see also \citealt{1999ApJ...520L..21M}).  They claim
that the origin of this bias is the excess of soft X-ray emission due
to small clumps of cool gas that continuously merge into the
intracluster medium.
 
In order to have a more realistic comparison with the spectroscopic
fits, a different definition of temperature was then introduced to
analyse the results of numerical simulations, namely the
emission-weighted estimate both including and excluding the
appropriate cooling function (see references in the next section).
However, even in this case there is a tendency to infer significantly
higher temperature values when the cluster has a complex thermal
structure (see, e.g., \citealt{2003astro.ph.10844G}).  In this
context, the main goal of our paper is to propose and test a new
formula for the theoretically estimated temperature.  Our formula
should make simulations more directly and accurately comparable to
observations.
 
The plan of the paper is as follows. In Section 2 we introduce the
problems originated by the projection of the cluster gas temperature
along the line of sight, both from an observational and a theoretical
point of view: the different definitions of projected temperature used
in the literature are here introduced. In Section 3 we discuss in
detail the effects of fitting realistic spectra - produced by
multi-temperature thermal models - by single-temperature models,
considering the specific case of \chandra ~ observations.  In Section 4
we propose a new analytic formula capable of approximating the cluster
spectroscopic temperatures measured from \chandra ~ and \xmm ~
observations with an accuracy always better than few percent. In
Section 5 we test the performance of this new relation by using the
outputs of a high-resolution hydrodynamical simulation, and compare
them to those obtained adopting the widely-used emission-weighted
estimate.  Finally in Section 6 we discuss our results and summarize
our conclusions.

\section{Projecting cluster gas temperatures along the line of  
sight: the problem}\label{par:problem} 
 
Although the atmosphere of galaxy clusters and groups is often far
from isothermal, it is common practice to assume for it a single
temperature, refereed as projected cluster temperature, in order to
identify the thermal structure of the system along the line of sight.
This is routinely done both by X-ray observers and by theoreticians,
who provide projected temperature estimates of real and simulated
clusters, respectively.  In the following we review the techniques
used to produce projected temperatures from simulated clusters and
discuss why they may differ from the estimates obtained through the
spectral analysis of observed clusters.
  
\subsection{Temperature from hydro-N-body simulations:  
mass- and emission-weighted temperatures}\label{sec:temp_def} 
 
Hydrodynamical N-body simulations provide information on the density
and temperature of each gas element (be it particle or grid point)
within the simulation box.  Generally speaking, workers in the field
of numerical simulations derive projected temperatures by simply
calculating the mean weighted value of the gas temperature along the
line of sight:
\be 
T\equiv \frac{\int W T dV} {\int W dV}\ , 
\label{eq:weight} 
\ee 
where $T$ is the temperature of a gas element, $dV$ is the volume
along the line of sight and $W$ is the weight function.
 
In literature there are essentially two classes of temperature
projection: mass-weighted and emission-weighted.  The first uses the
mass of the gas element as weighting function (i.e. $W\equiv m$; see,
e.g., \citealt{1994ApJ...428....1K}; \citealt{1996MNRAS.283..431B};
\citealt{2001ApJ...546..100M}):
\be 
T_{\rm mw}\equiv \frac{\int m T dV} {\int m dV}\ .
\ee 
This temperature definition was first introduced because of its
relevant physical meaning: the total thermal energy of the gas is
simply $E\propto m T_{\rm mw}$.  Despite of this, there has always
been a general concern on the fact that such definition would give
temperature estimates which differ significantly from what an X-ray
observer would derive via spectral analysis.  The main reason for this
discrepancy is that the X-ray emissivity $\epsilon_E$ is proportional
to the square of the gas density ($\epsilon_E \propto n^2$ rather than
$\propto n$), so it is expected that the spectroscopic temperature,
based on the number of emitted photons, is determined more by regions
at higher density than by those at lower density.  In order to make
the temperature estimates of simulated clusters more similar to the
spectroscopic one, a new kind of temperature was introduced, named
emission-weighted, where the weighting function is proportional to the
emissivity of each gas element ($W\equiv \epsilon_E$):
\be 
T_{\rm ew}\equiv \frac{\int \Lambda(T) n^2  T dV} {\int \Lambda(T) n^2 dV}\ , 
\ee 
with $\Lambda(T)$ the cooling function and $n$ the gas density  
(see, e.g., \citealt{1995MNRAS.275..720N}). 
 
Most of the works in the literature use the so called bolometric
cooling function $\Lambda(T)=\int_0^\infty \epsilon_E dE \propto \sqrt
{T}$ (see, e.g., \citealt{1998ApJ...495...80B};
\citealt{1999ApJ...525..554F}; \citealt{2001ApJ...552L..27M}), 
implicitly assuming bremsstrahlung (free-free) emission, which is the
dominant mechanism at temperatures larger than 3 keV.
 
If not otherwise specified, all the emission-weighted temperature
estimates reported in this paper will refer to this definition. In
other theoretical papers a much simpler definition is used, that
assumes $\Lambda(T)=1$ (see, e.g., \citealt{1994ApJ...428....1K};
\citealt{2001ApJ...546..100M}). This
definition is also known as {\em emission-measure-weighted}
temperature.  Only recently some authors (see, e.g.,
\citealt{2004MNRAS.348.1078B}) 
tried to improve the correspondence between simulations and
observations made with specific X-ray observatories, by adopting the
cooling function integrated on the specific telescope energy band:
$\Lambda(T)=\int_{E_{\rm min}}^{E_{\rm max}} \epsilon_E dE$, rather
than the bolometric one.
 
In the following we will show that none of these temperature
definitions are accurate approximations of the observed spectroscopic
temperature.
  
\subsection{Spectroscopic projected temperature} 
\label{par:spec_prop} 

From the viewpoint of an X-ray observation, the cluster gas
temperature is obtained by a fit of a thermal model to the observed
spectrum.  Measuring a projected temperature is thus equivalent to
finding a thermal model with temperature $T_{\rm spec}$ whose spectral
properties are as close as possible to the properties of the projected
spectrum.
 
Now, from plasma physics we know that, if the emitting gas is a single
temperature thermal plasma, its spectrum can be written as a linear
combination of continuum and line emission processes:
\be 
\epsilon_{E}=\epsilon_{E}^{\rm cont}+\epsilon_{E}^{\rm line}, 
\ee 
where $E$ is the photon energy.  For high gas temperatures ($T>3$~keV)
and/or low metallicity ($Z<<1$, with $Z$ in units of solar
metallicity) the continuum emission dominates over the line emission.
Let us assume, for now, that plasma metallicity is zero, so that we
can set $\epsilon_{E}^{\rm line}=0$.  In this case the emission
spectrum can be written as:
\be 
\epsilon_{E}^{\rm cont}\propto n_e^2G_c(Z,T,E)\frac{1}{\sqrt T}  
\exp\left(-\frac{E}{kT}\right),  
\label{eq:cont} 
\ee  
where $n_e$ is the gas electron density.  The function $G_c(Z,T,E)$ is
the total Gaunt factor, which is the sum of the Gaunt factors of three
main different continuum emission processes, namely free-free ($ff$,
thermal bremsstrahlung), free-bound ($fb$), and two-photon
($2\gamma$).
 
Regardless of the functional form of the Gaunt factor, it is
self-evident that, from a purely analytic point of view, the total
spectrum induced by two thermally isolated plasmas with electron
density $n_1$ and $n_2$ and different temperature $T_1$ and $T_2$ can
no longer be described by a single-temperature thermal model, with any
temperature $T_3$.  In fact,
\be 
\begin{array}{l} 
n_{1}^2G_c(Z,T_1,E)\frac{1}{\sqrt T_1}\exp(-\frac{E}{kT_1})+\\ 
n_{2}^2G_c(Z,T_2,E)\frac{1}{\sqrt T_2}\exp(-\frac{E}{kT_2}) 
\neq n_e^2G_c(Z,T_3,E)\frac{1}{\sqrt T_3}\exp(-\frac{E}{kT_3}),\\ 
\end{array} 
\ee  
unless $T_1=T_2$.  This consideration is quite interesting, as it
tells us that in principle the spectroscopic projected temperature is
not at all a well defined quantity.  In fact, as the spectrum of any
single-temperature model cannot completely reproduce the spectral
properties of a multi-temperature source, the inferred spectroscopic
temperature is a quantity that in principle may depend on the
efficiency of the X-ray detector used for the observation and on the
energy band used to fit the spectrum.  This is extremely important, as
it tells us immediately that a lot of attention must be paid when we
compare observational temperatures with the emission-weighted
temperatures defined earlier in \S~\ref{sec:temp_def}.  A proper
comparison between simulations and observations requires the actual
simulation of the spectral properties of the clusters via an X-ray
observatory simulator, like, for example, X-MAS
\citep{2003astro.ph.10844G}.
  
%
%
 
\begin{figure*} 
\psfig{figure=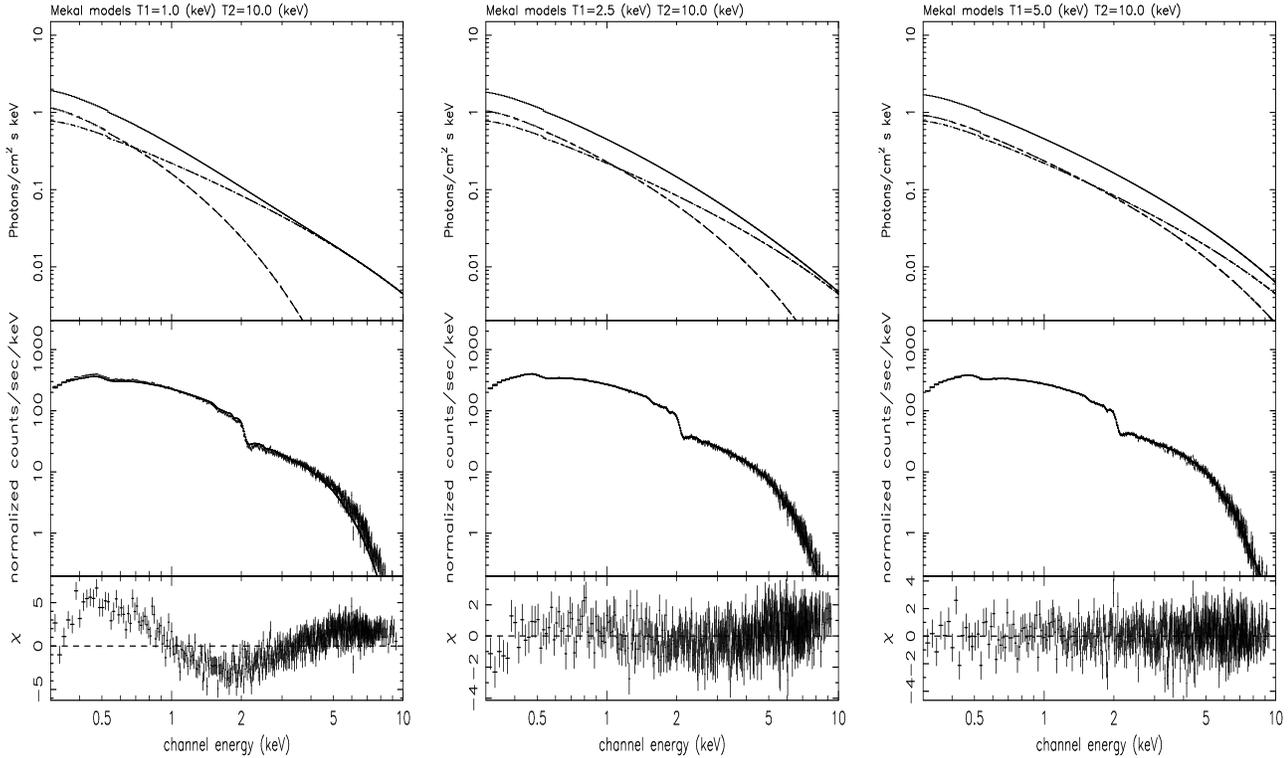,height=10.cm,width=17cm,angle=0} 
\caption{Upper panels:  
spectra of the input absorbed two-temperature thermal \textsc{mekal}
model (solid line).  The higher temperature model (dot-dashed line) is
$T_2=10$~keV in all panels.  The lower temperature model (dashed line)
is $T_1=1$~keV, $T_1=2.5$~keV, and $T_1=5$~keV for the left, middle,
and right panels, respectively.  In all panels we use
$n_H=10^{20}~{\rm cm}^{-2}$ and $Z=0$.  Middle panels: simulated
\chandra ~ spectra corresponding to the input model in the upper panels and
best fit model with one single thermal \textsc{mekal} model. 
Lower panels: residuals (in units  of $\sigma$) with error bars of size
one.}
\label{fig:spectra} 
\end{figure*} 
 
%
%
 
\begin{figure*} 
{\centering \leavevmode   
\psfig{file=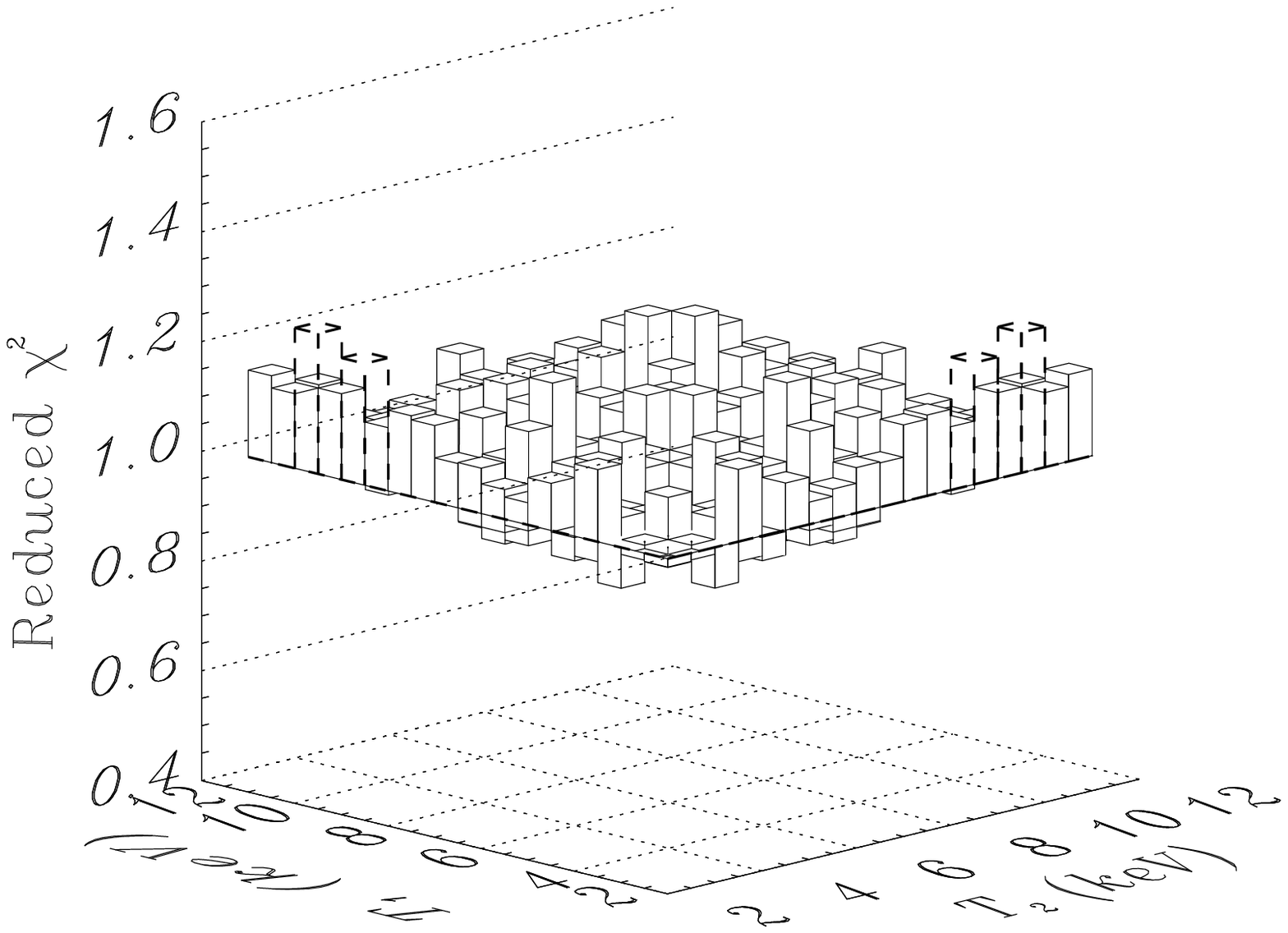,width=.49\textwidth} \hfil   
\psfig{file=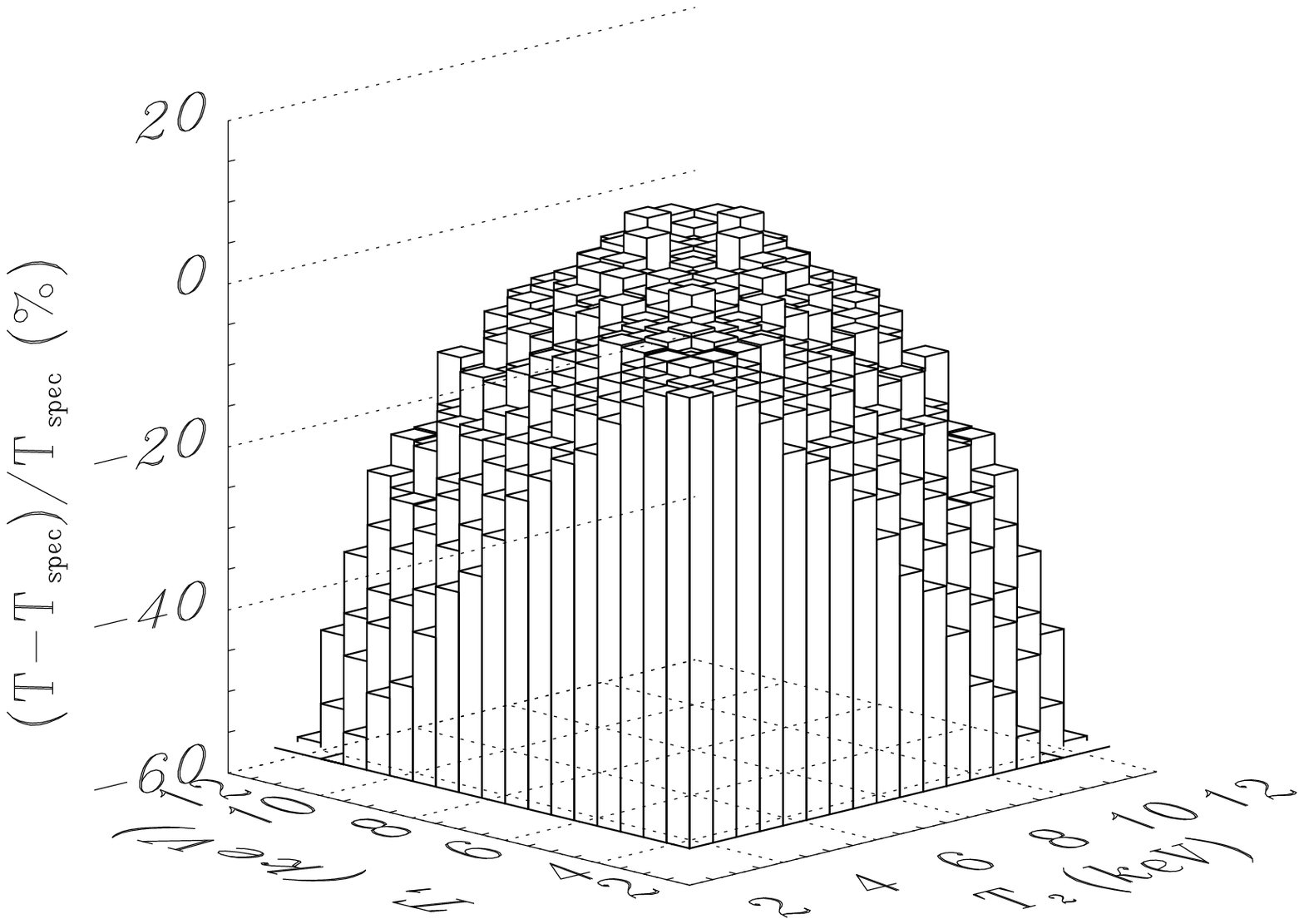,width=.49\textwidth}  
}  
\caption{Left panel: reduced $\chi^2$ corresponding to 
the best fit model of a single-temperature thermal model to the
simulated spectra of a two-temperature thermal model in the case
$Z=0$.  Each spectral fit has $\approx 500$ d.o.f. Solid and dashed
histograms indicate the fits for which the null hypothesis probability
is larger and smaller than 5 per cent, respectively. Right panel:
percentile difference between the calculated ($T$) emission-weighted
and spectroscopic (\tspec ) temperatures as a function of the
temperatures of the lower and higher model components ($T_1$ and
$T_2$, respectively).}
\label{fig:T_comp} 
\end{figure*} 

\section{Spectroscopic Projected Temperature from \chandra ~ Observations} 
\label{par:spec_proj} 
 
In the previous section we argued that from a purely analytic point of
view the spectrum of a multi-temperature thermal model cannot be
reproduced by any single temperature thermal model.  In the real world
things may be a bit different as, even assuming the most favorable
conditions, observed spectra are affected by at least the following
factors: i) convolution with the instrument response; ii) Poisson
noise; iii) instrumental and cosmic backgrounds.  These factors all
conjure to distort and, at some level, confuse the observed spectra.
Consequently it can happen that, under some circumstances, observed
spectra produced by multi-temperature thermal sources may be well
fitted by single-temperature thermal models which have little to do
with the real temperature, but nevertheless are statistically
indistinguishable from it. In the present section we want to address
exactly this issue.  In order to do that, we generated a number of
multi-temperature spectra and performed the standard fitting procedure
using a single-temperature model.
 
To simulate the spectra we used the command \textsc{fakeit} in the
utility XSPEC (\citealt{1996adass...5...17A}; see, e.g., Xspec User's
Guide version 11.2.x; \citealt{2001adass..10..415D}
\footnote{http://legacy.gsfc.nasa.gov/docs/xanadu/xspec/manual}).
This command creates simulated data from the input spectral model by
convolving it with the ancillary response files (ARF) and the
redistribution matrix files (RMF), which fully define the response of
the considered instrument, and by adding the noise appropriate to the
specified integration time.  To begin with, in the following
subsections we will concentrate on observations made with \chandra ~ in
ACIS-S configuration.  In particular, all simulated spectra will be
obtained by using the ARF and RMF relative to the aim point of the
chip ACIS-S3.  Later, we will discuss the differences between ACIS-S3
and the ACIS-I \chandra ~ and MOS and pn \xmm ~ detectors.  In order
to make the simulated spectra more similar to the actual observations,
we multiply each spectrum by an absorption model (\textsc{wabs} in
XSPEC) that accounts for the galactic absorption.  In this paragraph
we fix the equivalent hydrogen column density of the absorption model to
$n_H=10^{20}~{\rm cm}^{-2}$.
 
Our main interest is in the possible spectral distortion induced by
instrument response, rather than in the spectral uncertainties
connected with the data statistics.  For this reason we will assume
that the observed spectra have no background, and will rescale
exposure times so that the total number counts per spectrum is very
large ($N\approx 350,000$).
 
All simulated spectra presented in this section are fitted by an
absorbed single-temperature \textsc{mekal} model using the $\chi^2$
statistic, and leaving $n_H$, $Z$, and $T$ as free
parameters. Spectral fits are done in the $0.3-10$~keV energy band.
Regardless of the quality of the fit, from now on we will call
spectroscopic temperature $T_{\rm spec}$ the best fit temperature
value resulting from the above fitting procedure.
 
%
%
 
\begin{figure*} 
\psfig{figure=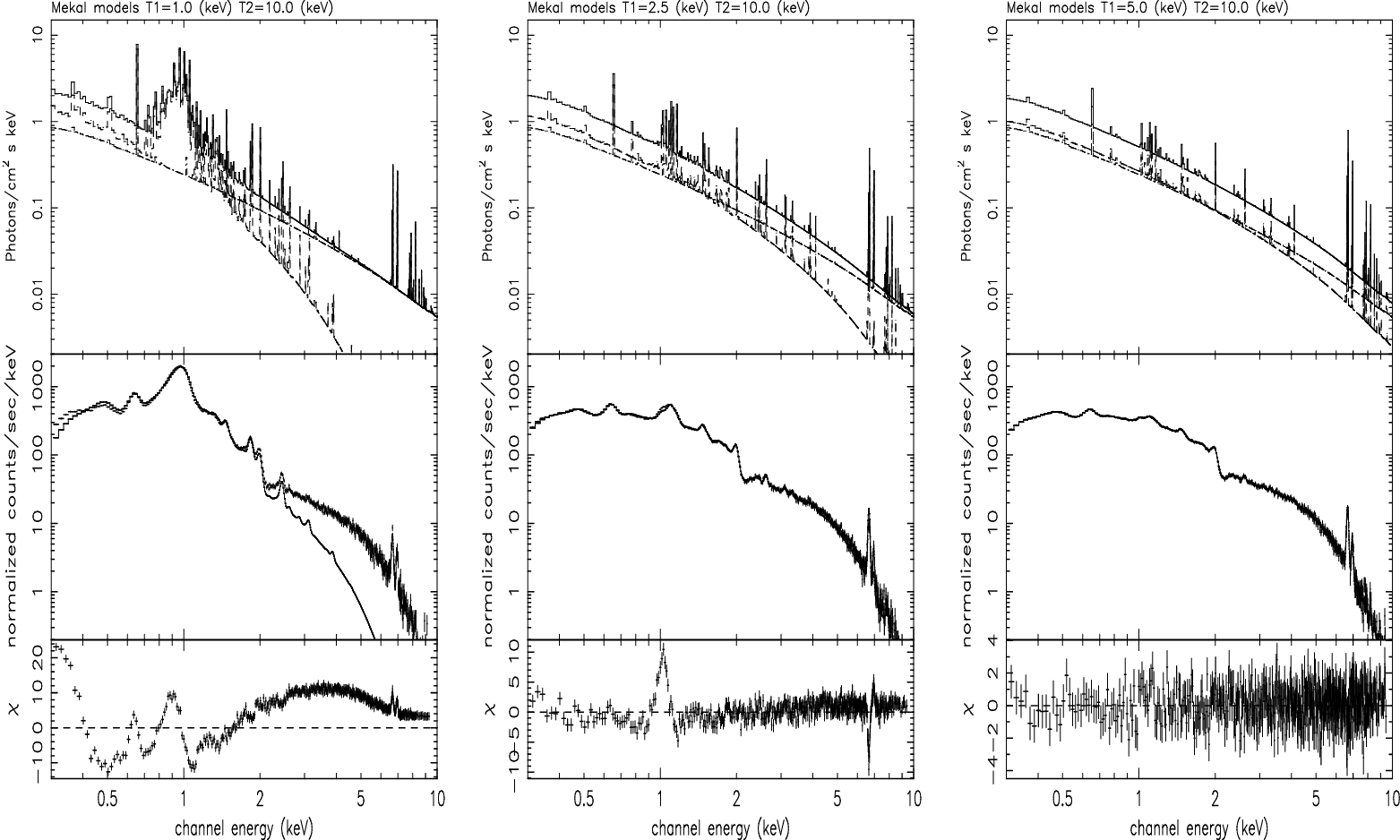,height=10.cm,width=17cm,angle=0} 
\caption{As Fig.~\ref{fig:spectra}, but for metallicity $Z=1$.}
\label{fig:spectra_1.0} 
\end{figure*} 
 
%
%
 
\begin{figure*} 
{\centering \leavevmode   
\psfig{file=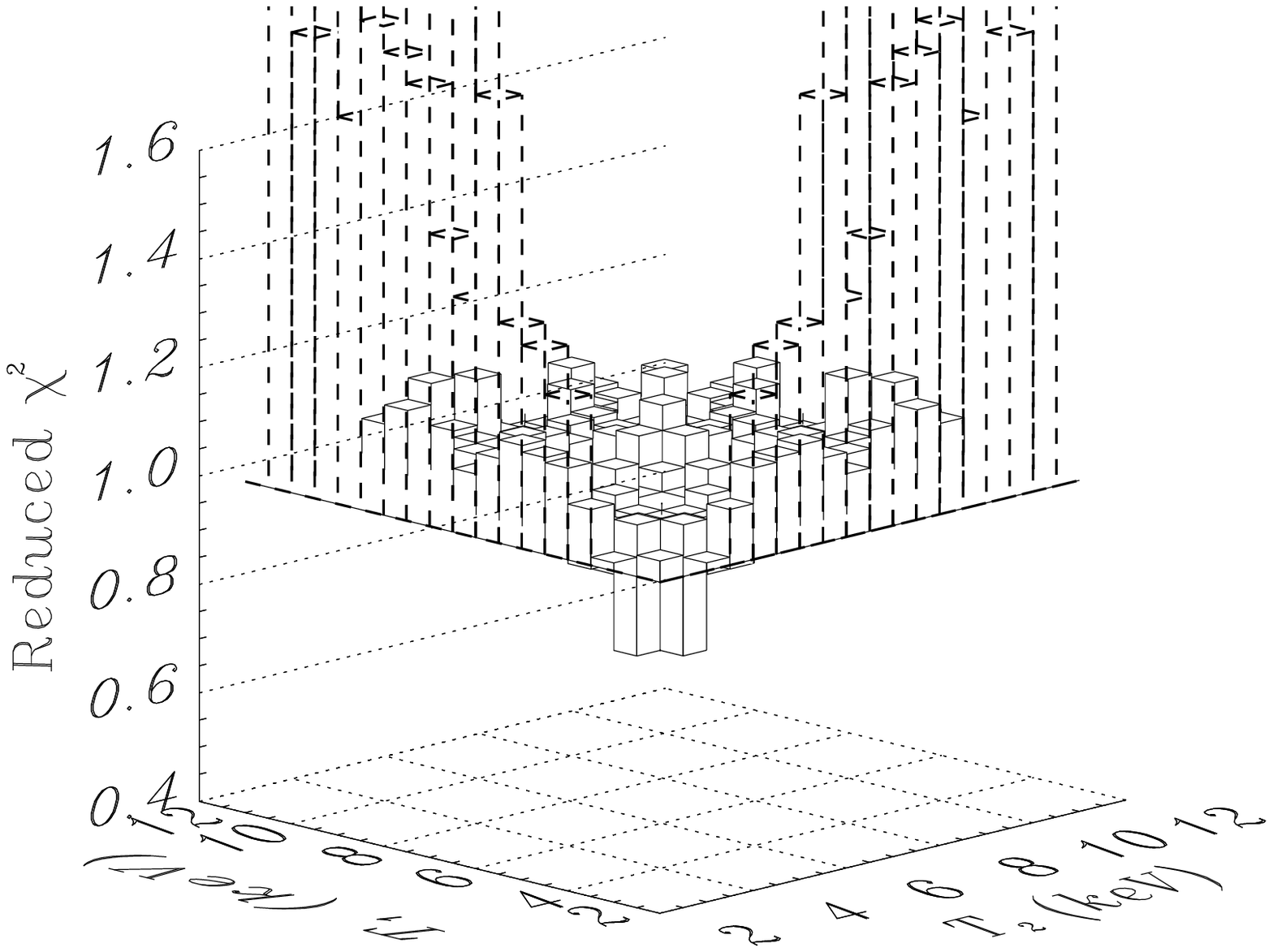,width=.49\textwidth} \hfil   
\psfig{file=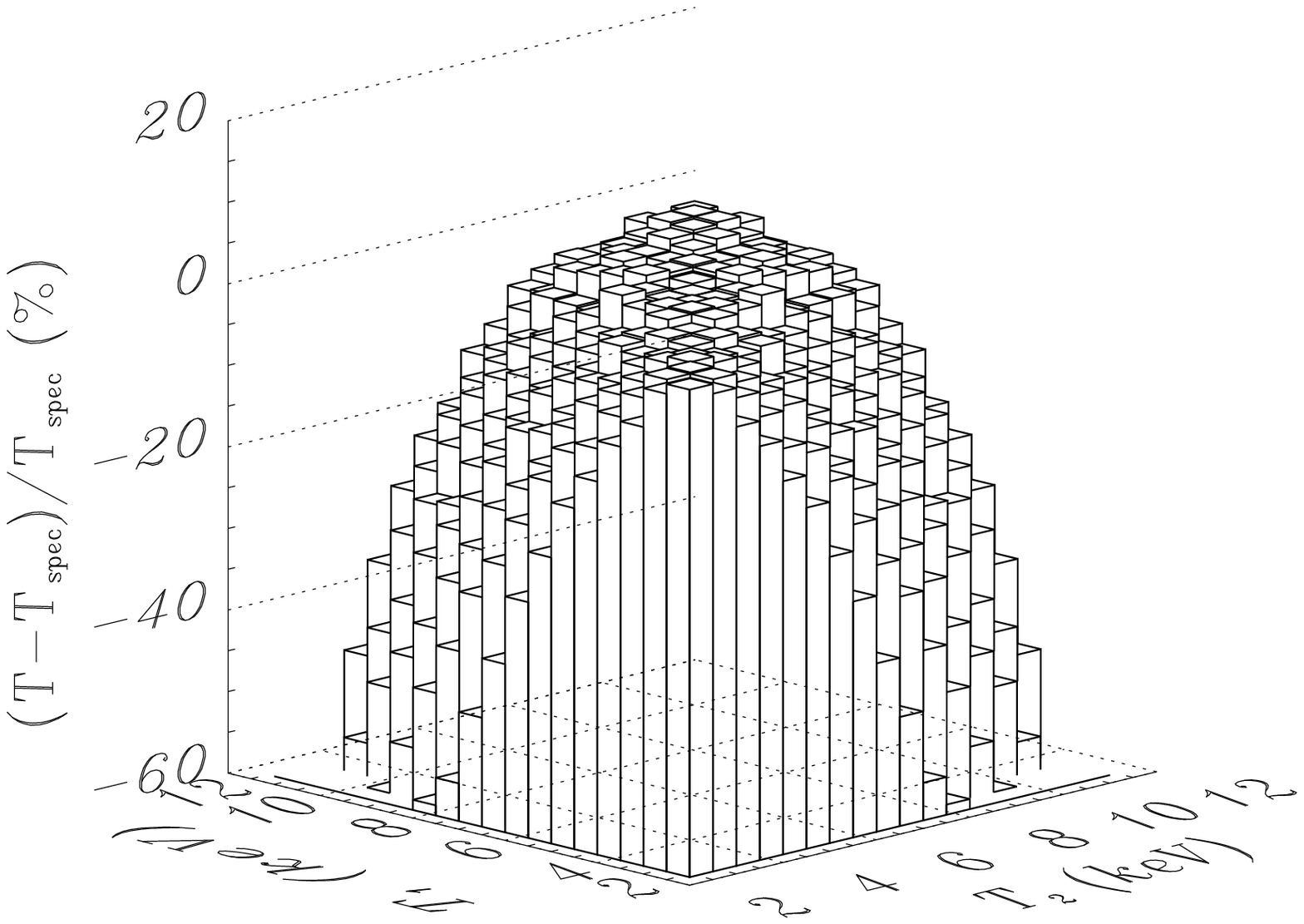,width=.49\textwidth}  
}  
\caption{As Fig.~\ref{fig:T_comp}, but for metallicity $Z=1$.} 
\label{fig:T_comp_1.0} 
\end{figure*} 
 
\subsection{Fitting two-temperature thermal spectra with   
single-temperature models: the zero-metallicity case}\label{par:z=0}
 
We will start by considering the simplest possible model, namely the
one in which the gas metallicity is $Z=0$.  In this case the source
emission is given by the sum of the continuum spectra (see
Eq.~\ref{eq:cont}) relative to each plasma component with different
temperatures.  For didactical purposes we first compare the simulated
spectrum of a two-temperature thermal source with the best fit from a
single-temperature model.  We will assume that the two components of
the thermal source have the same emission measure ($k\propto \int
n_e^2 dV$).  Let us consider three examples, named spectrum 1, 2, and
3 for convenience.  All spectra are characterized by the same higher
temperature component, fixed at $T_2=10$~keV.  The lower temperature
component is instead $T_1=1$~keV, $T_1=2.5$~keV, and $T_1=5$~keV for
spectrum 1, 2, and 3, respectively.  In the left, middle, and right
panels of Fig.~\ref{fig:spectra} we report the two-temperature thermal
input models, the simulated source spectrum, the best fit of the
single-temperature thermal model, and the residuals in units of the
temperature dispersion $\sigma$ for spectrum 1, 2, and 3,
respectively.  The figure clearly shows that the source spectrum~1
($T_1=1$~keV) cannot be fitted by a single-temperature thermal model.
Conversely, spectrum~3 ($T_1=5$~keV) is well fitted by a single
temperature thermal model and actually is statistically
indistinguishable from it.  It is interesting to note that also
spectra with low-temperature components $T_1< 5$~keV can actually be
fitted quite reasonably by a single-temperature model: in fact,
although spectrum~2 ($T_1=2.5$~keV) shows some small departures from
its best fit single-temperature model (see the middle panels of
Fig.~\ref{fig:spectra}), the $\chi^2$/d.o.f=526/513 indicates that the
fit is still statistically acceptable.
 
Starting from the three examples above we can explore the intervals of
$T_1$ and $T_2$ required for the source spectra to be well fitted by a
single-temperature model.  To do so, we produced a large set of
two-temperature source spectra with different $T_1$ and $T_2$, and
estimated the spectroscopic temperature for each.
 
The first interesting result we find is the following: consistently
with what found by \cite{2001ApJ...546..100M}, whenever the lower
temperature component is at $T>2$~keV, then almost any two-temperature
spectrum can be well fitted by a single-temperature model.  This is
clearly shown on the left panel of Fig.~\ref{fig:T_comp}, where we
report the reduced $\chi^2$ relative to the best fit model of a
single-temperature thermal model to the simulated spectra of a
two-temperature thermal model.  The different line styles indicate if
the fit is statistically acceptable or not.  In particular, solid and
dashed histograms refer to the fits for which the null hypothesis has
a probability $> 5$ per cent (statistically acceptable fit) or $<5$
per cent (statistically unacceptable fit), respectively.  We also
notice that the reduced $\chi^2$ is always very close to unity, except
in few cases where the lower temperature component is at $T \sim
2$~keV and the higher temperature component is at $T>8$~keV.  It is
important to say that, for lower temperature components at $T<2$~keV,
our results show that a single-temperature model is generally not a
good fit, unless the difference $|T_2-T_1|$ is small ($<0.5$~keV). In
particular, the lower the temperature of the low-temperature
component, the smaller should be difference $|T_2-T_1|$ in order for
its spectrum to be compatible with a single-temperature model.
 
A second important result we find is that the spectroscopic
temperature $T_{\rm spec}$ is always lower than the bolometric
emission-weighted temperature, in agreement with reports by some
authors (see, e.g., \citealt{2001ApJ...546..100M};
\citealt{2003astro.ph.10844G}).  This is clearly shown on the right
panel of Fig.~\ref{fig:T_comp}, where we plot the percentile
difference between the emission-weighted and spectroscopic
temperatures as a function of the lower and higher temperature
components of the model: we notice that the larger $|T_2-T_1|$, the
larger this difference, with values as large as 60 per cent for
$T_2<12$~keV that can be even larger for higher $T_2$.
 
%
%
 
\begin{figure} 
{\centering \leavevmode 
\psfig{file=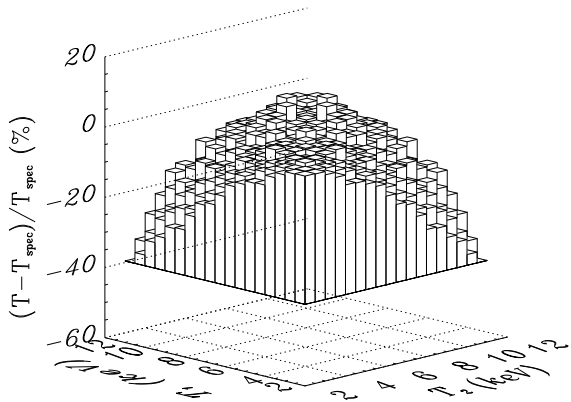,width=.45\textwidth} \hfil 
} 
\caption{Percentile difference between the calculated ($T$)  
emission-measure-weighted $T$ (with weight $W\propto n_e^2$) and the
spectroscopic ($T_{\rm spec}$) temperatures as a function of the lower
and higher model component temperatures ($T_1$ and $T_2$,
respectively).}
\label{fig:chi2} 
\end{figure} 
  
\subsection{Fitting two-temperature thermal spectra with   
single-temperature models: the case $Z=1$}\label{par:z=1}
 
We now study the effect of the line emission on the spectroscopic
temperature estimates.  To do that we repeat the analysis of
\S~\ref{par:z=1}, but now assuming $Z=1$.
 
As done in the previous subsection, we first show the spectra for the
three example models, namely spectra 1, 2, and 3.  On the left,
middle, and right panels of Fig.~\ref{fig:spectra_1.0} we report the
two-temperature thermal input models, the simulated source spectrum,
the best-fit single-temperature thermal model, and the residuals in
terms of $\sigma$ for spectrum 1, 2, and 3, respectively.  By
comparing Fig.~\ref{fig:spectra} with Fig.~\ref{fig:spectra_1.0} we
notice that the presence of metals in the plasma induces the formation
of lines that may not be reproduced by the best-fit single-temperature
model, as clearly shown in the middle panels of
Fig.~\ref{fig:spectra_1.0}.  Although the continuum of the source
spectrum is very well reproduced by the best fit single-temperature
model (see. Fig.~\ref{fig:spectra}), the line emission is
under-predicted at $1$~keV and over-predicted at $7$~keV (see residual
plot in Fig.~\ref{fig:spectra_1.0}). This suggests that
single-temperature models are not suited to fit multi-temperature
spectra.  Nevertheless, when the low-temperature component is high
enough, then a single-temperature model can still fit very well a
two-temperature source spectrum, and again it becomes statistically
indistinguishable from it (right panels of
Fig.~\ref{fig:spectra_1.0}). This is because the higher the
temperatures, the less important the line contribution to the final
spectrum; in such cases the source with $Z=1$ becomes more and more
similar to that with $Z=0$.
 
As done in \S~\ref{par:z=0}, we produced a set of synthetic
two-temperature source spectra with different $T_1$ and $T_2$, and
estimated for each the spectroscopic temperature.  In the left panel
of Fig.~\ref{fig:T_comp_1.0} we report the reduced $\chi^2$ relative
to the best fit single-temperature model. As expected, we find that in
this case the number of source spectra that cannot be fitted by a
single-temperature model is higher.  Nevertheless, all source spectra
whose lower temperature component is smaller than 3 keV continue to be
statistically indistinguishable from a single-temperature model.
 
Is is worth noticing that also in the case with $Z=1$ we find that the
spectroscopic temperature is always lower than the emission weighted
one, as clearly shown in the right panel of Fig.~\ref{fig:T_comp_1.0}.
We also notice that the observed percentile difference in temperature
is higher compared with the case with $Z=0$, although only by a few
per cent.
 
We also tested the goodness-of-fit of a much simpler
emission-measure-weighted temperature function, i.e. the one obtained
by assuming $\Lambda(T)=1$ (see \S~\ref{sec:temp_def}).  The
percentile difference between the spectroscopic and
emission-measure-weighted temperature calculated in this way is shown
in Fig.~\ref{fig:chi2}.  Paradoxically, we find that this simpler
temperature definition fares better than the other one, even if it
still overpredicts $T_{\rm spec}$ with differences that, for
$T_2<12$~keV, can be as large as 40 per cent.  We must add that a
different definition of $T_{\rm ew}$, which uses the cooling function
integrated in the telescope energy band
($\Lambda(T)=\int_{E_{\rm min}}^{E_{\rm max}} \epsilon_E dE$), gives results
in-between the two discussed here.  These results demonstrate that
none of the emission-weighted temperature functions so far used in the
literature actually provides a good approximation of the projected
spectroscopic temperature obtained directly from observations.
  
\begin{figure*} 
{\centering \leavevmode   
\psfig{file=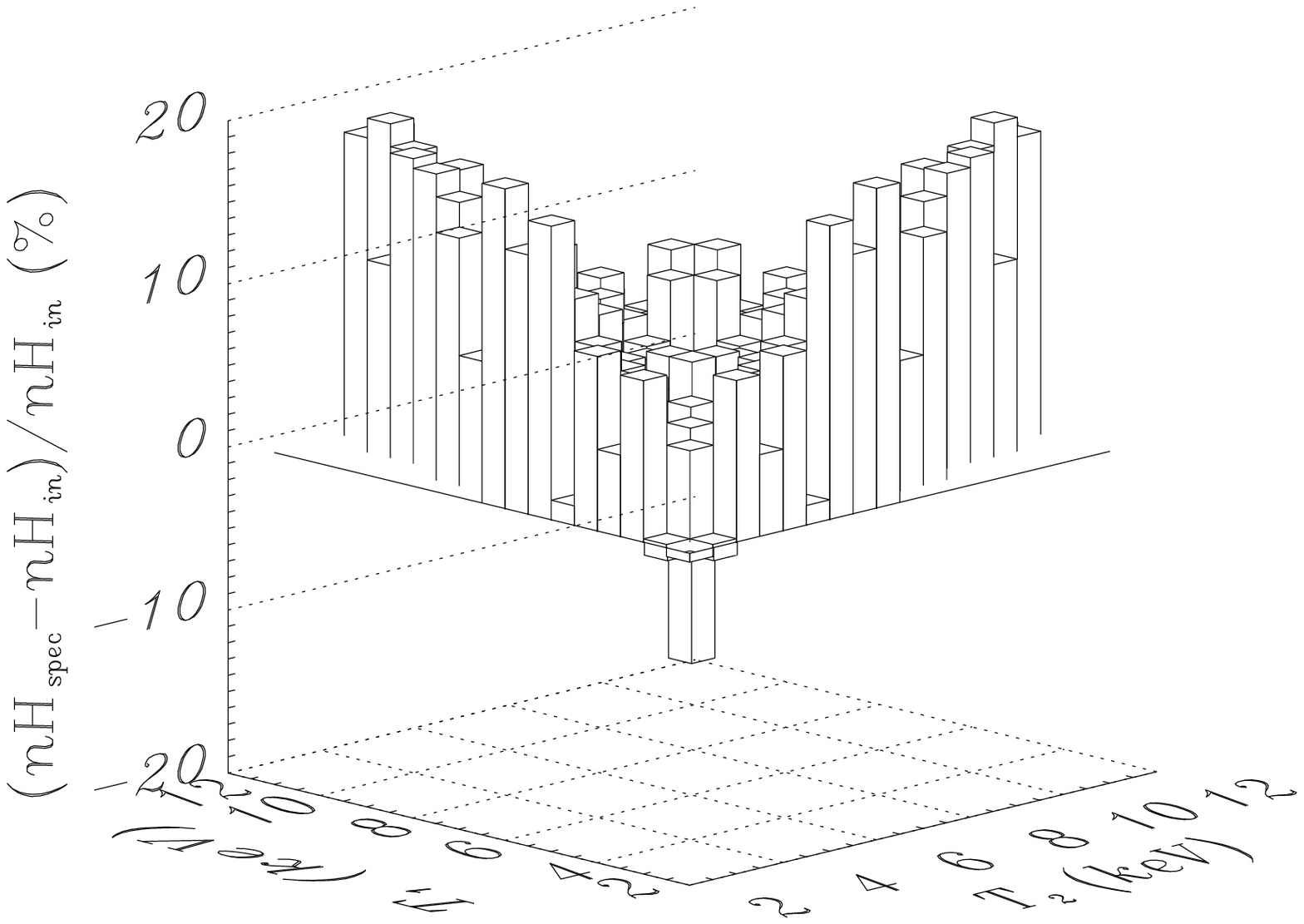,width=.49\textwidth} \hfil   
\psfig{file=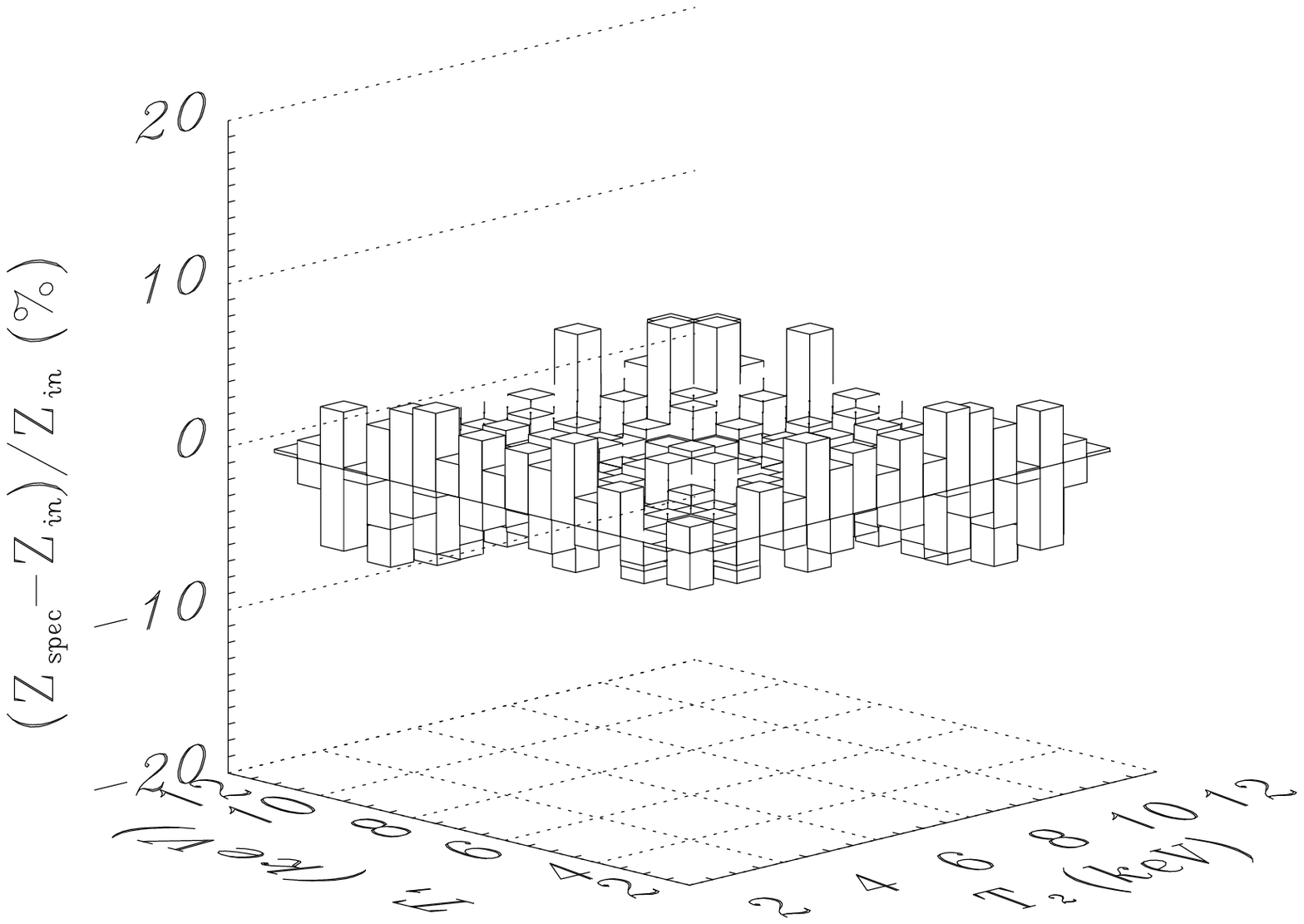,width=.49\textwidth}  
}  
\caption{ 
Percentile difference between the value obtained by fitting a
single-temperature model to a two-temperature source spectra and the
input value.  Left and right panels refer to the equivalent hydrogen
column density $n_H$ (input value $n_{H_{\rm in}}=10^{20}{\rm
cm}^{-2}$) and to the metallicity $Z$ in solar units (input value
$Z_{\rm in}=1$).}
\label{fig:n_h_1.0} 
\end{figure*}

For completeness, we conclude this section by reporting the effects of
projection on the hydrogen column density and metallicity estimates.
To do that we compare the fitted $n_H$ and $Z$ of the
single-temperature model with the input values.  Percentile
differences are shown in the left and right panels of
Fig.~\ref{fig:n_h_1.0}, respectively. Interestingly, we find that both
the measured $n_H$ and $Z$ are different from the input values. This
difference, however, is always within 10 per cent\footnote{Notice: the
histogram in the right panel of Fig.~\ref{fig:n_h_1.0} where the
percentile variation of $n_H$ is smaller than 10 per cent corresponds
to regions where the fit is not statistically acceptable.}.
 
%
%
 
%
%
 
\begin{figure*} 
{\centering \leavevmode   
\psfig{file=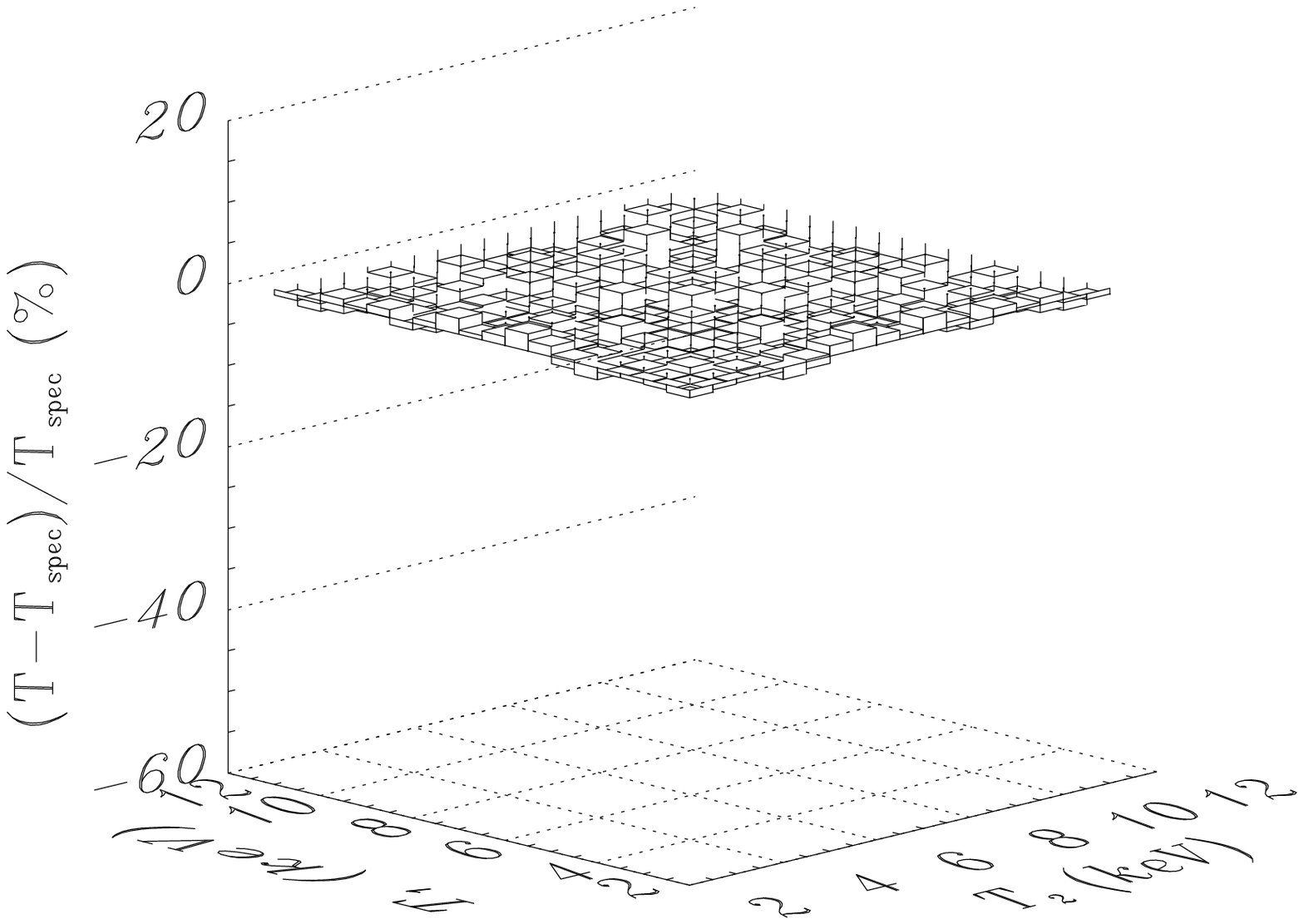,width=.49\textwidth} \hfil 
\psfig{file=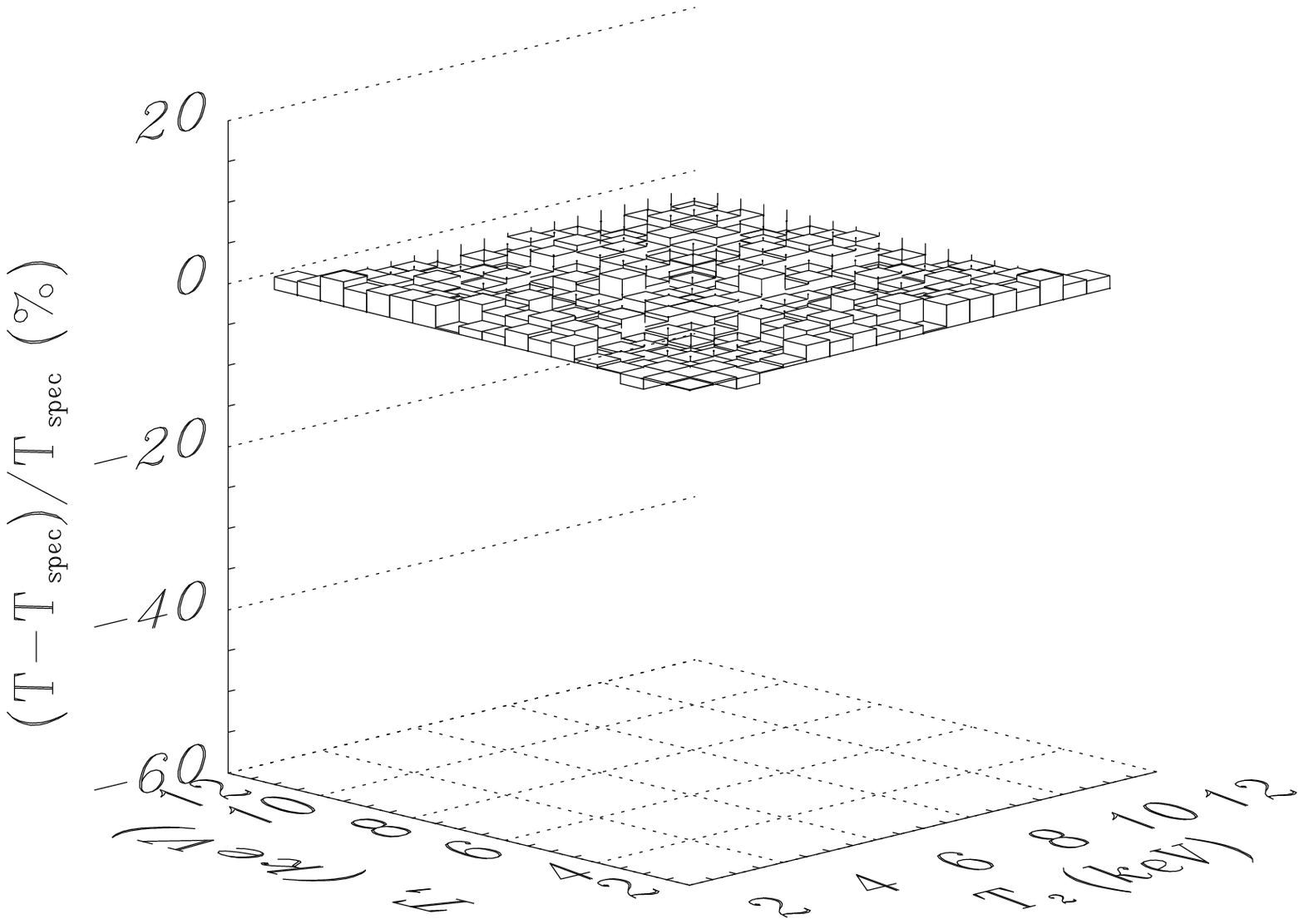,width=.49\textwidth} 
}  
\caption{Percentile difference between the calculated  ($T$) and the 
spectroscopic ($T_{\rm spec}$) temperature a a function of the lower
and higher model component temperatures ($T_1$ and $T_2$,
respectively). To calculate $T$ we use the spectroscopic-like
temperature $T_{\rm sl}$ defined in Eq.~\ref{eq:new_form-fin} with the
value of $\alpha$ chosen in order to minimize the value of
$\Delta$. Left panel: observational conditions defined as in
\S~\ref{par:z=0} (to be compared with the right panel of
Fig.~\ref{fig:T_comp}). Right panel: observational conditions defined
as in \S~\ref{par:z=1} (to be compared with the right panel of
Fig.~\ref{fig:T_comp_1.0}).}
\label{fig:T_comp_new} 
\end{figure*} 
  
\section{Approximate formula to estimate  \chandra ~ and  
\xmm ~ spectroscopic temperatures}\label{par:slike} 
 
In the previous section we have shown that, under some circumstances,
a multi-temperature source spectrum can be fitted by a
single-temperature model, providing the estimate $T_{\rm spec}$.  We
have also demonstrated that $T_{\rm spec}$ is always lower than the
bolometric emission-weighted temperature $T_{\rm ew}$.  In this
section we wish to derive a projected emission-weighted temperature
formula that can better approximate the spectroscopic one.

The idea is quite simple: given a multi-temperature thermal emission
we want to identify the one temperature whose spectrum is closest to
the observed spectrum.  From now on, we will call $T_{\rm sl}$ this
"spectroscopic-like" temperature.  If we assume two thermal components
with constant densities $n_1$, $n_2$, 
and temperatures $T_1$, $T_2$, respectively, requiring matching
spectra means that
\be 
\begin{array}{l} 
n_{1}^2\zeta(Z,T_1)\frac{1}{\sqrt T_1}\exp(-\frac{E}{kT_1})+ 
n_{2}^2\zeta(Z,T_2,)\frac{1}{\sqrt T_2}\exp(-\frac{E}{kT_2})\\ 
\approx A\zeta(Z,T_{\rm sl})\frac{1}{\sqrt T_{\rm 
    sl}}\exp(-\frac{E}{kT_{\rm sl}}),\\ 
\end{array} 
\label{eq:condizion} 
\ee  
where $A$ is an arbitrary normalization constant and $\zeta(Z,T)$ is a
parametrization function that accounts for the total Gaunt factor and
partly for the line emission.
 
Both \chandra ~ and the \xmm ~ are most sensitive to the soft region
of the X-ray spectrum, so we can expand both sides of
Eq.~\ref{eq:condizion} in Taylor series, to the first order in $E/kT$:
\be 
\begin{array}{l} 
n_{1}^2\zeta(Z,T_1)\frac{1}{\sqrt T_1}(1-\frac{E}{kT_1})+ 
n_{2}^2\zeta(Z,T_2)\frac{1}{\sqrt T_2}(1-\frac{E}{kT_2})\\ 
\approx A\zeta(Z,T_{\rm sl})\frac{1}{\sqrt T_{\rm 
    sl}}(1-\frac{E}{kT_{\rm sl}}).\\ 
\end{array} 
\label{eq:condizion_taylor} 
\ee  
 
By equating the zero-th and first-order terms in $E$, we finally find
the equation defining the spectroscopic-like temperature:
\be 
T_{\rm sl}\approx 
\frac{n_1^2\zeta(Z,T_{1})/T_1^{1/2}+n_2^2\zeta(Z,T_{2})/T_2^{1/2}} 
{n_1^2\zeta(Z,T_{1})/T_1^{3/2}+n_2^2\zeta(Z,T_{2})/T_2^{3/2}}. 
\label{eq:new_form-2} 
\ee  
 
The extension of Eq.~\ref{eq:new_form-2} to a continuum distribution
of plasma temperatures in a volume $V$ is trivial:
\be 
T_{\rm sl}\approx \frac{\int n^2\zeta(Z,T)/T^{1/2} dV } 
{\int n^2\zeta(Z,T)/T^{3/2} dV}. 
\label{eq:new_form-n} 
\ee  
 
To calculate the function $\zeta(Z,T)$ we consider the fact that it
gives, together with $n^2$, the relative contribution of the spectral
component with temperature $T$ to the total spectrum.  The amplitude
of this normalization contribution is set partly by the total Gaunt
factor, partly by the line emission. In principle, their relative
importance depends on the details of the instrument/analysis used
(e.g. the energy band used for the fit) and on the observational
conditions (e.g. low or high galactic absorption regions).  However,
for temperatures and metallicities realistic in clusters of galaxies
($0<Z<1$, $0.5~{\rm keV}<T<20~{\rm keV}$) the total Gaunt factor
depends essentially on the temperature, and can be approximated by a
power-law relation:
\be 
G_c(Z,T,E)\approx (T/{\rm keV})^{\eta}, 
\label{eq:gaunt} 
\ee 
where $\eta\approx 0.25$ (see, e.g., \citealt{1981A&AS...45...11M}).
The effect of the lines is to increase the plasma emissivity,
especially at low energies: this effect can also be approximated by a
power-law of the temperature. Thus in the following we will assume for
$\zeta(Z,T)$ the following functional form:
\be 
\zeta(Z,T)\propto (T/{\rm keV})^{\alpha},\\ 
\label{eq:zeta1} 
\ee 
where the parameter $\alpha$ depends on the specific observational
conditions and on the used instrument, so it needs to be appropriately
calibrated.  We do this by adopting the following procedure: we select
some specific observation conditions, simulate a set of
two-temperature source spectra as explained in \S~\ref {par:spec_proj}
and calculate the mean value of the percentile variation of $T_{\rm
sl}$ with respect to $T_{\rm spec}$: \be
\Delta=\frac{1}{N}\sum_{T_1,T_2} \sqrt{\Big ( \frac {T_{\rm sl}-T_{\rm 
      spec}}{T_{\rm spec}}\Big)^2}. 
\label{eq:find-min} 
\ee 
In the previous formula the sum is extended to all the $N$ simulated
source spectra with $3~{\rm keV} <T_1< 20~{\rm keV}$ and $3~{\rm keV}
<T_2< 20~{\rm keV}$.  The variable $\alpha$ is obtained though a
minimization procedure of $\Delta$.
 
\begin{table} 
\scriptsize 
\centering 
\caption{Value of $\alpha$ and corresponding minimum $\Delta$ for different 
  observational conditions}
\label{table:1} 
\begin{tabular}{ c c c c c c c }  
\hline \hline 
       Number & Detector & $n_H$ & $Z$&$E_{\rm min}$ & $\alpha$
       &minimum $\Delta$\\ & & ($10^{22}$cm$^{-2}$) & &(keV) &&\\
\hline 
1                      & ACIS-S3  &   0                                      &0                &0.3 &    0.73     &    2.1\\ 
2                      & ACIS-S3  &   1                                      &0                &0.3 &    0.86     &    3.7\\ 
3                      & ACIS-S3  &   0                                      &1                &0.3 &    0.41     &    1.4\\ 
4                      & ACIS-S3  &   1                                      &1                &0.3 &    0.63    &    2.3 \\ 
&&&& &&\\                                                                                                                    
5                      & ACIS-S3  &   0                                      &0                &1.0 &    0.79     &    3.0\\ 
6                      & ACIS-S3  &   1                                      &0                &1.0 &    0.89     &    3.9\\ 
7                      & ACIS-S3  &   0                                      &1                &1.0 &    0.48     &    2.1\\ 
8                      & ACIS-S3  &   1                                      &1                &1.0 &    0.82     &    2.5\\ 
\hline 
 
1                      & MOS  &   0                                      &0                &0.3  &    0.82   &     2.3\\ 
2                      & MOS  &   1                                      &0                &0.3  &    0.92   &     3.7\\ 
3                      & MOS  &   0                                      &1                &0.3  &    0.46   &     1.5\\ 
4                      & MOS  &   1                                      &1                &0.3  &    0.81   &     2.3\\ 
&&&& &&\\                                                                                                              
5                      & MOS  &   0                                      &0                &1.0  &    0.86   &     3.4\\ 
6                      & MOS  &   1                                      &0                &1.0  &    0.91   &     4.0\\ 
7                      & MOS  &   0                                      &1                &1.0  &    0.55   &     2.2\\ 
8                      & MOS  &   1                                      &1                &1.0  &    0.96   &     2.4\\ 
\hline                                                                                           
\end{tabular} 
\end{table} 

To show how well $T_{\rm sl}$ reproduces the actual $T_{\rm spec}$, we
apply it to the cases discussed earlier in \S~\ref{par:z=0} and
\S~\ref{par:z=1}. The minimization procedure of $\Delta$ returns
$\alpha=0.86$ and $\alpha=0.65$ for the Z=0 and Z=1, respectively.
The resulting percentile variation between $T_{\rm sl}$ and $T_{\rm
spec}$ for these two cases is shown in Fig.~\ref{fig:T_comp_new}, on
the left and right panels, respectively.  The figure clearly shows
that, unlike $T_{\rm ew}$, $T_{\rm sl}$ provides a good estimate of
the observed spectroscopic temperature, at a level better than 2-3 per
cent.
  
\begin{figure} 
{\centering \leavevmode   
\psfig{file=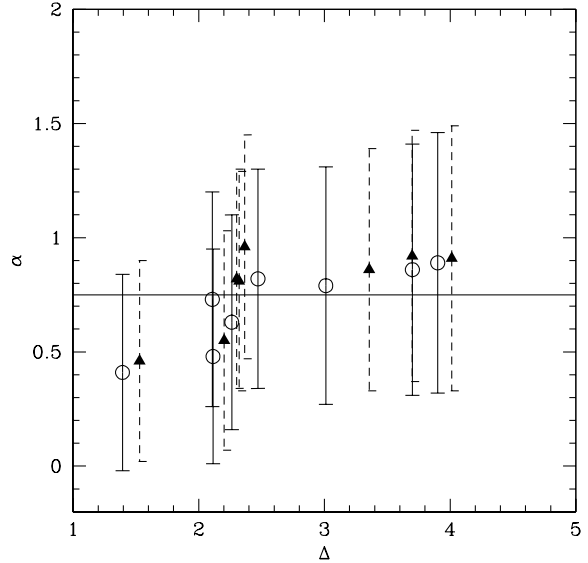,width=.45\textwidth}  
}  
\caption{Value of $\alpha$  and corresponding minimum value of $\Delta$ 
for the observational conditions reported in Table~\ref{table:1}.
Open circles and filled triangles refer to \chandra ~ ACIS-S3 and \xmm
~ MOS, respectively. The error bars show the range in $\alpha$ where
$\Delta<5 $ per cent (see text). The horizontal line indicates
$\alpha=0.75$. }
\label{fig:alpha-delta} 
\end{figure}

We now study how the power index $\alpha$ depends on the observational
conditions for all the \chandra ~ and \xmm ~ detectors, and how it
changes as a function of the minimum energy $E_{\rm min}$ considered
in the spectral fitting.  In Table~\ref{table:1} we report for some of
the cases considered the value of $\alpha$ obtained using the
minimization procedure, and the corresponding value of minimum
$\Delta$.  The cases reported in the table are only a few examples of
those we actually considered. They represent extreme situations for
galaxy clusters, in the sense that we explicitly considered strong and
null galactic absorption ($n_H=10^{22}~{\rm cm}^{-2}$ and $n_H=0~{\rm
cm}^{-2}$, respectively), and solar (Z=1) and zero (Z=0) metal
abundances.  The values of $\alpha$ and the corresponding minimum
$\Delta$ from Table~\ref{table:1} are also shown in
Fig.~\ref{fig:alpha-delta}.  Open circles and filled triangles refer
to \chandra ~ ACIS-S3 and \xmm ~ MOS, respectively.  From
Fig.~\ref{fig:alpha-delta} it is clear that, as expected, the value of
$\alpha$ minimizing the discrepancy between $T_{\rm sl}$ and $T_{\rm
spec}$ depends on the actual observation conditions.  Nevertheless, we
notice that: i) $\alpha$ and $\Delta$ are very close to each other for
\chandra ~ and \xmm ; ii) the minimum $\Delta$ is generally very low
and smaller than 4 per cent.  Following these considerations we
explored the interval in $\alpha$ for which the corresponding value of
$\Delta$ is smaller than 5 per cent. The results are reported in
Fig.~\ref{fig:alpha-delta}, where this range is shown as y-axis error
bars.  We notice that all the error bars overlap. This is important as
it means that, regardless of the observational conditions, we can
select a value of $\alpha$ (which we call $\alpha_u$) so that $T_{\rm
sl}$ can reproduce $T_{\rm spec}$ with an accuracy better than $5$ per
cent on average.  As shown by the solid horizontal line in
Fig.~\ref{fig:alpha-delta} a good choice for \chandra ~ and \xmm ~ is
$\alpha_u=0.75$.
  
%
%
 
%
%
  
\begin{figure*} 
{\centering \leavevmode   
\psfig{file=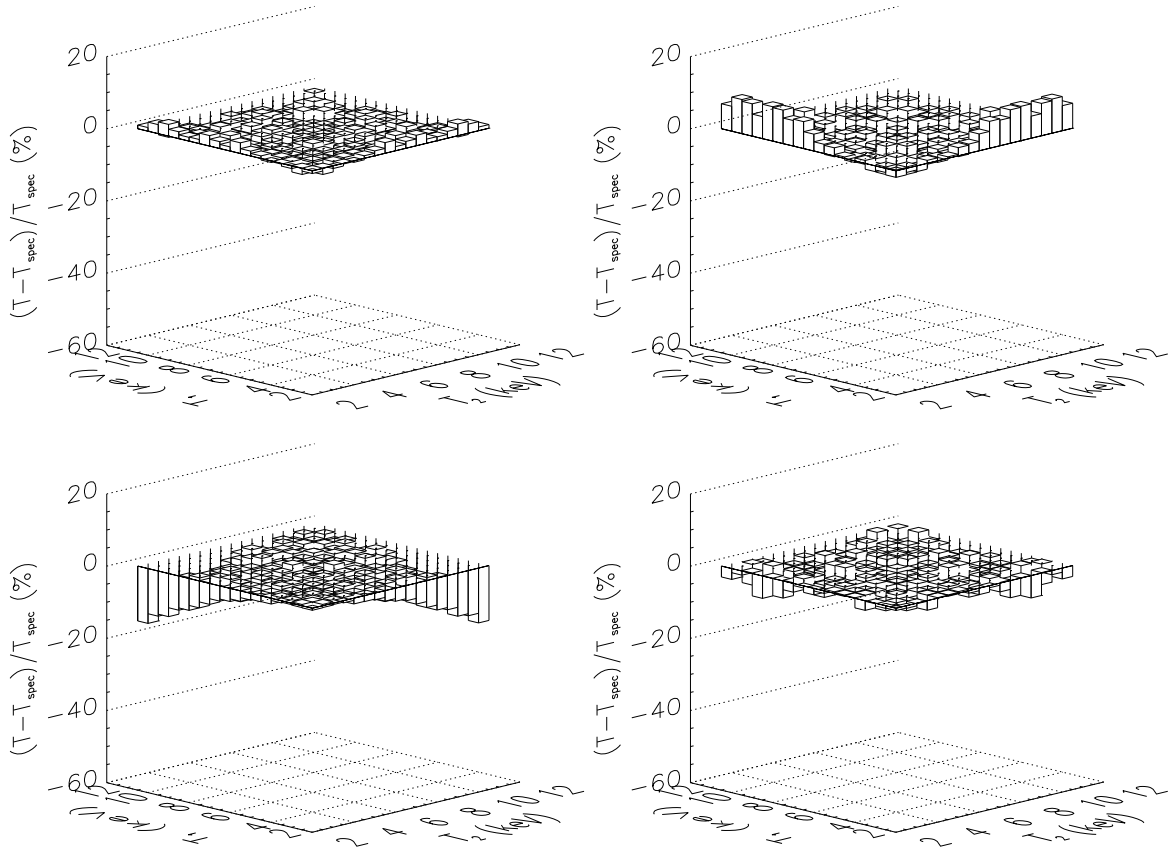,width=.9\textwidth}  
}  
\caption{Percentile difference between the calculated  ($T$) and the 
spectroscopic ($T_{\rm spec}$) temperatures as a function of the lower
and higher model component temperatures ($T_1$ and $T_2$,
respectively). To calculate $T$ we use the spectroscopic-like
temperature $T_{\rm sl}$ defined in Eq.~\ref{eq:new_form-fin} assuming
$\alpha=\alpha_u\equiv 0.75$.  The panels from up to down, left to
right, correspond to the observational conditions 1 to 4, reported in
Table~\ref{table:1} with ACIS-S3, respectively. }
\label{fig:t_t_sl:0.3} 
\end{figure*} 
 
\begin{figure*} 
{\centering \leavevmode   
\psfig{file=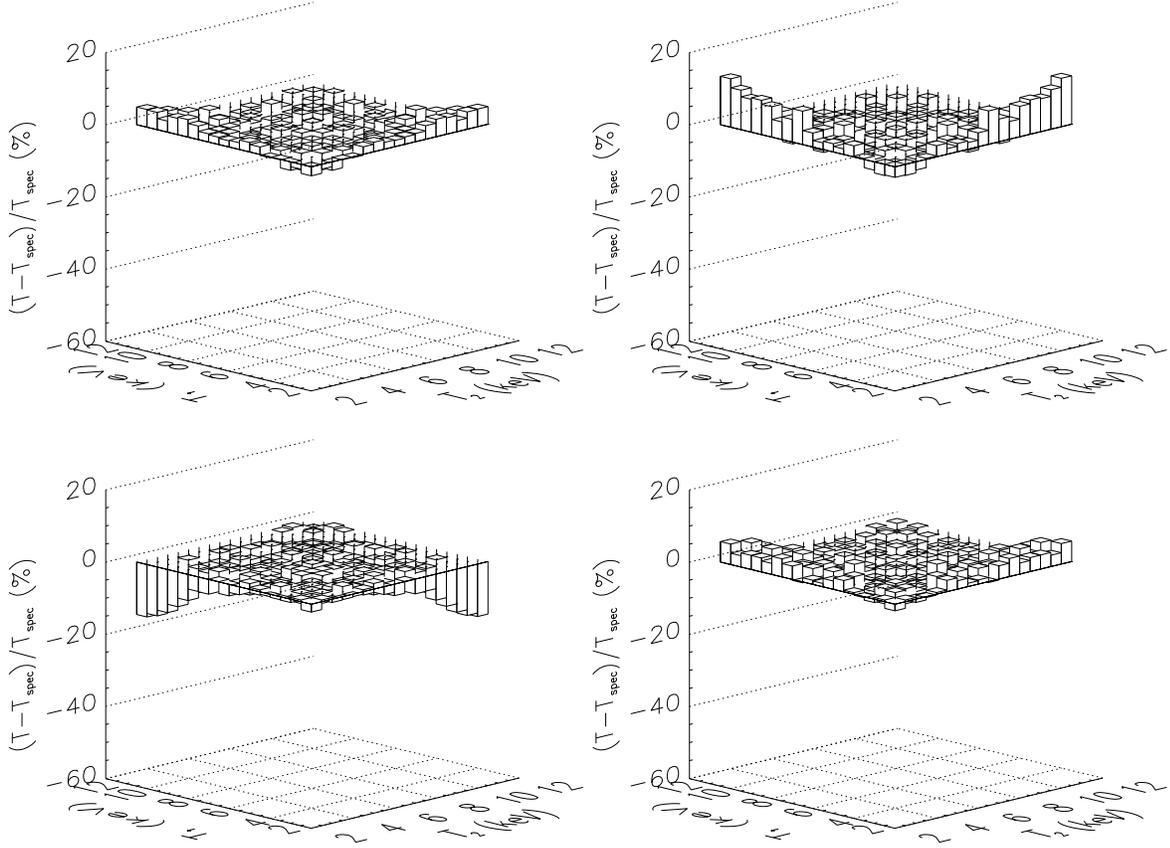,width=.9\textwidth}  
}  
\caption{As Fig.~\ref{fig:t_t_sl:0.3}, but the panels from up to 
down, left to right, correspond now to the observational conditions 5
to 8 reported in Table~\ref{table:1} with ACIS-S3, respectively.  }
\label{fig:t_t_sl:1} 
\end{figure*}

The accuracy of $T_{\rm sl}$ when we adopt $\alpha_u=0.75$ is shown in
Fig.~\ref{fig:t_t_sl:0.3} and Fig.~\ref{fig:t_t_sl:1}, where we report
the percentile variation between the values of $T_{\rm sl}$ and
$T_{\rm spec}$ obtained in different observational conditions with the
\chandra ~ ACIS-S3 detector.  In Fig.~\ref{fig:t_t_sl:0.3} and
Fig.~\ref{fig:t_t_sl:1} the values of $T_{\rm spec}$ were obtained by
fitting the spectra down to $E_{\rm min}=0.3$~keV and $E_{\rm
min}=1$~keV, respectively.  The four panels in each figure correspond
to ($n_H=0\times 10^{22}$cm$^{-2}$, $Z=0$), ($n_H=1\times
10^{22}$cm$^{-2}$, $Z=0$), ($n_H=0\times 10^{22}$cm$^{-2}$, $Z=1$),
and ($n_H=1\times 10^{22}$cm$^{-2}$, $Z=1$).  As expected, for most of
the temperature combinations of the source spectra, $T_{\rm
sl}(\alpha_u=0.75)$ gives a good estimate of $T_{\rm spec}$ at a level
better than 5 per cent, while the difference is higher only for a few
spectra (see, e.g., upper-right and lower-left panels of
Fig.~\ref{fig:t_t_sl:0.3} and Fig.~\ref{fig:t_t_sl:1}).  However, the
only spectra for which the discrepancy is higher than $5$ per cent are
those on the border of the temperature plane, i.e. those whose the
lower temperature component in the source model is $T=3$~keV.  As
already explained in \S~\ref{par:z=0} and \S~\ref{par:z=1}, this
temperature represents a sort of ``borderline'' value for \chandra ~
and \xmm . Thus, the reason of such higher discrepancy is simply
explained by the inability to properly define a spectroscopic
projected temperature for these spectra, due to the poor quality of
the fit with a single temperature model (see left panels of
Fig.~\ref{fig:T_comp} and \ref{fig:T_comp_1.0}).

%
%
 
To summarize the results of this section, we claim that the
spectroscopic-like temperature defined by:
\be 
T_{\rm sl}(\alpha)\approx \frac{\int n^2T^{\alpha}/T^{1/2} dV } 
{\int n^2T^{\alpha}/T^{3/2} dV}, 
\label{eq:new_form-fin} 
\ee  
with $\alpha=\alpha_u=0.75$, gives a good approximation of the
spectroscopic temperature $T_{\rm spec}$ obtained from data analysis
of \chandra ~ and \xmm ~ observations.
 
We notice that Eq.~\ref{eq:new_form-fin} can be rewritten in the
general form of Eq.~\ref{eq:weight}:
\be 
\begin{array}{rl} 
T_{\rm sl}&= \frac{\int W T dV }{\int W dV}, \\ 
W&=\frac{n^2}{T^{3/4}}.\\ 
\end{array} 
\label{eq:new_form_TW}  
\ee  
It is interesting to note that $T_{\rm sl}$ weights each thermal
component directly by the emission measure but, unlike \tew, inversely
by their temperature to the power of $3/4$.  This means that, beside
being biased toward the densest regions of the clusters, the observed
spectroscopic temperature is also biased toward the coolest
regions. In the next sections we will discuss some important
implications of this bias.
  
\section {Testing Spectroscopic-Like Temperature on Hydro-N-body 
simulation}\label{par:testing}
 
Using a simplified two-temperature thermal model we have shown that
the spectroscopic-like temperature $T_{\rm sl}$ provides a good
approximation to the spectroscopically derived temperature $T_{\rm
spec}$.  Here we wish to extend this study by showing the goodness of
\tsl ~ in reproducing \tspec ~ in a more general case.
 
In order to do that, we follow the same procedure used in
\cite{2003astro.ph.10844G}, where we use the output of an hydro-N-body
simulation and the X-ray MAp Simulator (X-MAS) to compare $T_{\rm
spec}$ with $T_{\rm ew}$ for a simulated cluster of galaxies. Here we
use the same simulation to extend the comparison to \tsl .
 
Let us briefly remind that the cluster simulation used in
\cite{2003astro.ph.10844G} was selected from a sample of 17 objects
obtained by re-simulating at higher resolution a patch of a
pre-existing cosmological simulation.  The assumed cosmological
framework is a cold dark matter model in a flat universe, with present
matter density parameter $\Omega_m=0.3$ and a contribution to the
density due to the cosmological constant $\Omega_\Lambda =0.7$; the
baryon content was set to $\Omega_B=0.03$; the value of the Hubble
constant (in units of 100 km/s/Mpc) is $h=0.7$, and the power spectrum
normalization is given by $\sigma_8=0.9$.  The re-simulation method we
used, called ZIC (for Zoomed Initial Conditions), is described in
detail in \cite{1997MNRAS.286..865T}, while an extended discussion of
the properties of the whole sample of these simulated clusters is
presented elsewhere (\citealt{2003astro.ph..4375T};
\citealt{2003astro.ph..9405R}).  Here we remind only some of the
characteristics of the cluster used in this paper.  The simulation was
obtained by using the publicly available code GADGET
(\citealt{2001NewA....6...79S}); during the run, starting at redshift
$z_{\rm in} = 35$, we took 51 snapshots equally spaced in $\log(1+z)$,
from $z=10$ to $z=0$.  The cluster virial mass at $z=0$ is $1.46
\times 10^{15} h^{-1} M_\odot$, corresponding to a virial radius of
$2.3h^{-1}$ Mpc; the mass resolution is $4.5\times 10^9 h^{-1}
M_\odot$ per dark particles and $5\times 10^8 h^{-1} M_\odot$ per gas
particles; the total number of particles found inside the virial
radius is 566,374, 48 per cent of which are gas particles.  The
gravitational softening is given by a $5h^{-1}$ kpc cubic spline
smoothing.
  
\begin{figure*} 
 {\centering \leavevmode 
\psfig{file=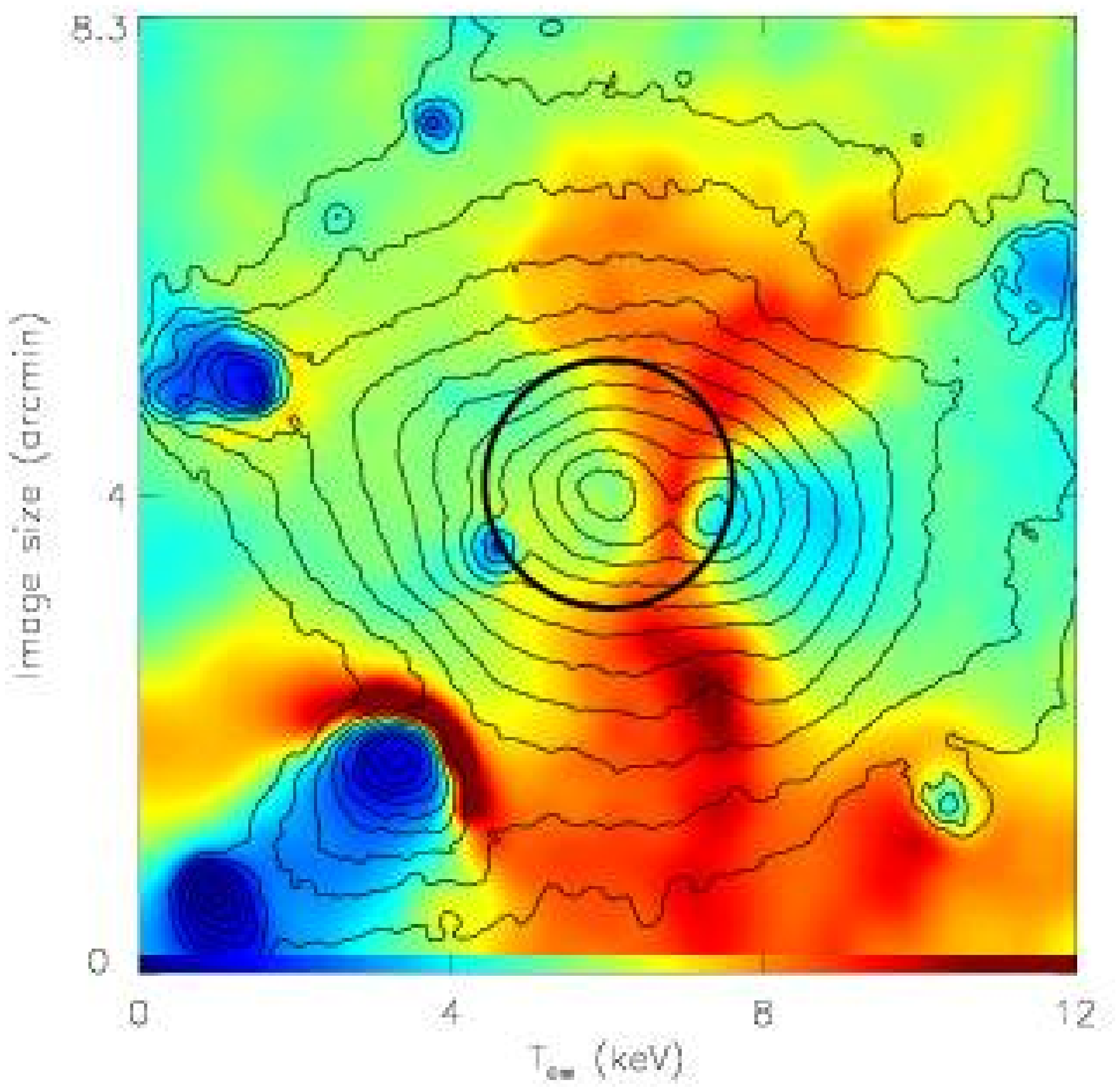,width=.45\textwidth}  
\psfig{file=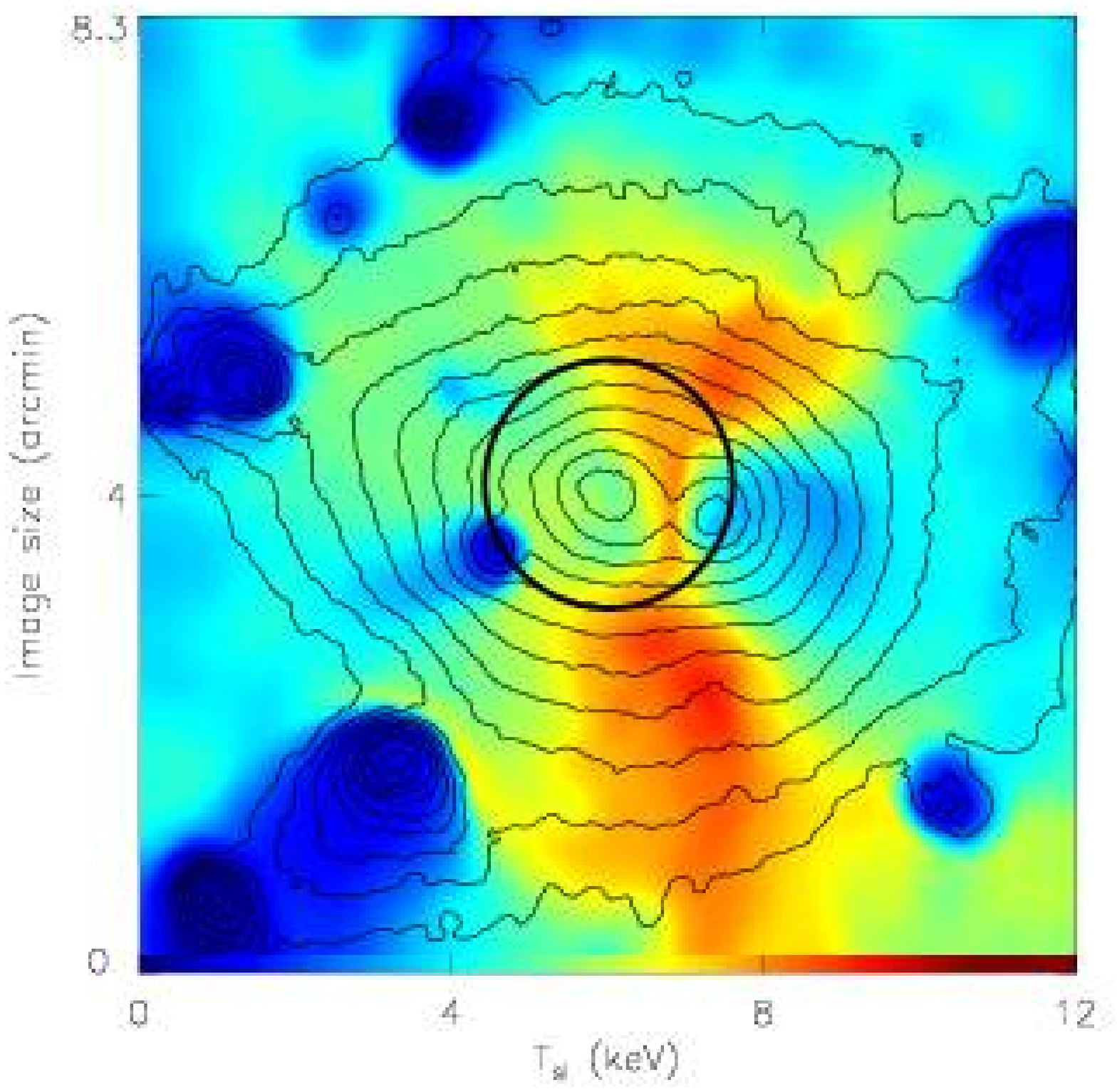,width=.45\textwidth}\hfil 
} 
\caption{Comparison of the map  of the 
emission-weighted temperature \tew ~ (left-panel) and the
spectroscopic-like temperature \tsl ~ (right-panel) for the simulated
cluster of galaxy.  Both maps are obtained by using the gas particles
of the hydro-N-body simulation and are binned in $1\arcsec$ pixels.
The contour levels correspond to the cluster flux distribution. The
circle shows the cluster region from which the temperature profile
sispayed in Fig.~\ref{fig:profiles} has been extracted.}
\label{fig:com} 
\end{figure*}

In the left panel of Fig.~\ref{fig:com} we report the
emission-weighted temperature map of the simulated clusters overlaid
to the cluster flux distribution from \cite{2003astro.ph.10844G}.  The
figure clearly shows that the cluster is far from isothermal.  Among
other features, we point to the reader the presence in this map of two
shock fronts.  The first is in the lower-left corner, and is produced
by the motion toward the cluster centre of the innermost of the two
subclumps present in that region. This front has a post-shock gas
temperature of approximately $18$~keV. A similar, but weaker, shock
front is located in front of the subclump, at the centre of the
Eastern side of the cluster.  For comparison, on the right panel of
Fig.~\ref{fig:com} we report the spectroscopic-like temperature map of
the same cluster, overlaid to the cluster flux distribution.  An
immediate thing to notice is that the map of $T_{\rm sl}$ appears
cooler than the map of $T_{\rm ew}$.  This is consistent with the fact
that, unlike $T_{\rm ew}$, $T_{\rm sl}$ is biased toward the lowest
values of the dominant temperature components along the line of sight.
Another very important point is that both shock fronts, which are
clearly evident in the emission-weighted temperature map, are no
longer detected in the $T_{\rm sl}$ map.  This aspect will be further
discussed in \S~\ref{par:dis} below.
 
In the following subsections we will compare $T_{\rm spec}$, $T_{\rm
ew}$ and $ T_{\rm sl}$ in two cases of practical interest: the
determinations of the projected cluster temperature map and the
determination of the radial profile.  To do that we will use the
300~ks \chandra ~ ACIS-S3 X-MAS ``observation'' of this cluster from
\citet{2003astro.ph.10844G}.  As explained in that paper, the
observation for this particular cluster was performed by fixing the
cluster metallicity to $Z=0.3\,Z_\odot$ and the column density to
$N_H=5\times 10^{20}$ cm$^{-2}$; data were analysed by applying the
standard procedures and tools used for real observations.  We refer to
\citet{2003astro.ph.10844G} for more details on the data analysis. 
Here we just remind that $T_{\rm spec}$ is obtained by fitting the
spectra in the [0.6,9.0]~keV energy band with a single/temperature
absorbed \textsc{mekal} model, after fixing the cluster redshift,
metallicity, and hydrogen column density to the values used as inputs
to compute the \chandra ~ observation.

%
%
 
\subsection {Cluster projected temperature map}\label{par:temap} 
  
\begin{figure} 
{\centering \leavevmode 
\psfig{file=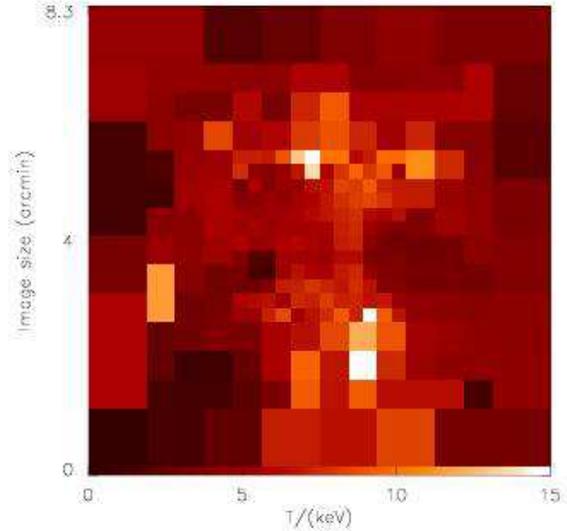,width=.45\textwidth}\hfil 
} 
\caption{Spectroscopic temperature map of  
the simulated cluster of galaxy as derived from the spectroscopic
analysis of the \chandra ~ ``observation'' with the package X-MAS
(from \citealt{2003astro.ph.10844G}).  }
\label{fig:comparison} 
\end{figure}

As mentioned above, we used the spectroscopic temperature map
published in \citet{2003astro.ph.10844G}.  That map was obtained by
subdividing the cluster image in squares large enough to contain at
least 250 net photons.  The projected spectroscopic temperature map
$T_{\rm spec}$ is shown in Fig.~\ref{fig:comparison}.  In order to
make a direct comparison of this map to those with emission-weighted
and the spectroscopic-like temperatures, we decreased the resolution
of the latter to match the resolution of the former.
 
\begin{figure*} 
{\centering \leavevmode 
\psfig{file=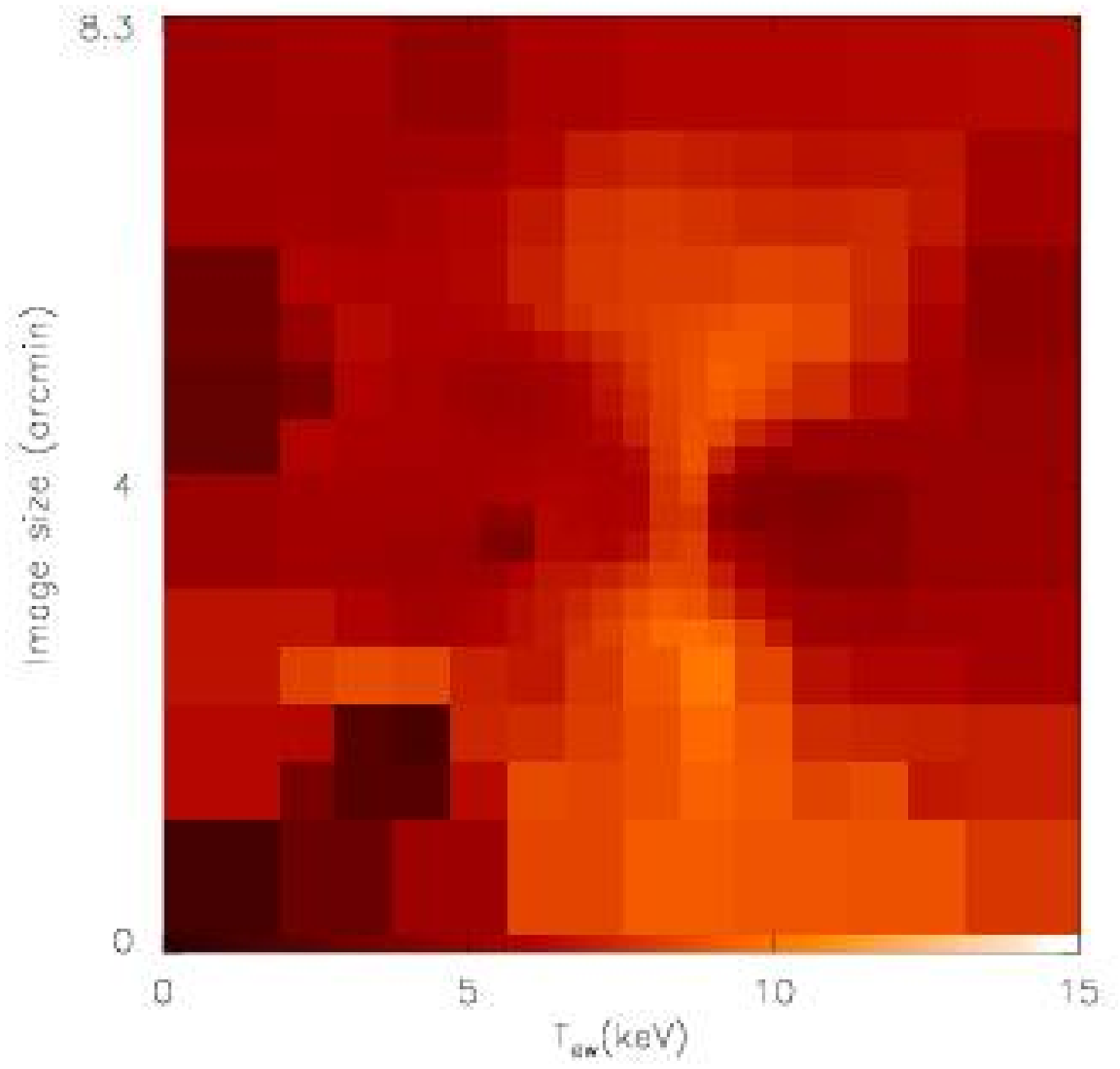,width=.45\textwidth} 
\psfig{file=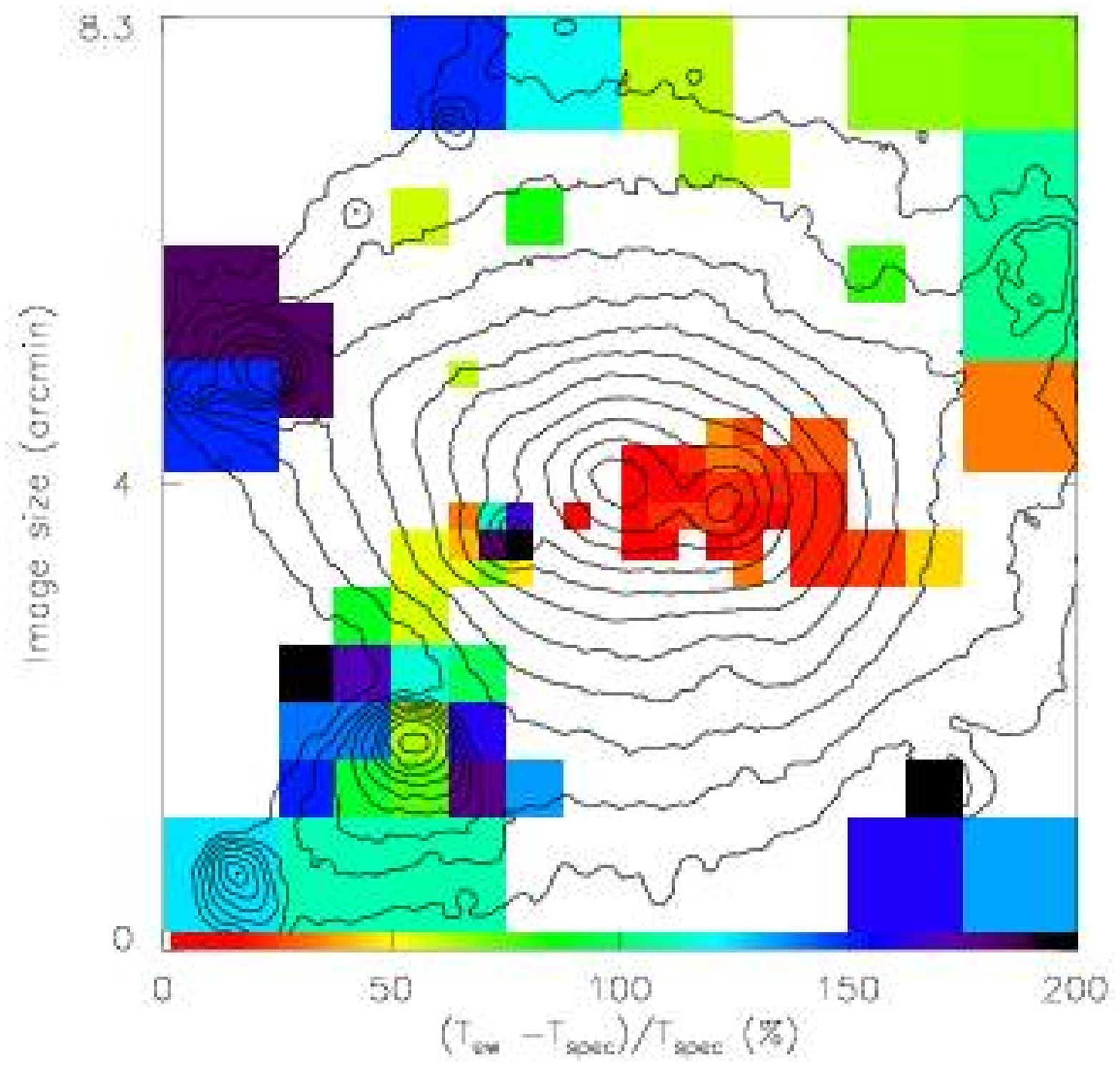,width=.45\textwidth} \hfil 
} 
\caption{Left panel: emission-weighted temperature map of the simulated 
cluster of galaxies shown in the left panel of Fig.~\ref{fig:com}
re-binned to match the spatial resolution of the spectroscopic
temperature map shown in Fig.~\ref{fig:comparison}.  Right panel:
percentile difference between the spectroscopic and emission-weighted
temperature maps.  In this map we show only the regions where the
significance level of the temperature discrepancy is at least $3
\sigma$, i.e.  $|(T_{\rm ew}-T_{\rm spec}) /\sigma_{\rm spec}|>3$,
where $\sigma_{\rm spec}$ is the 68 per cent confidence level error
associated to $T_{\rm spec}$.}
\label{fig:comparison2} 
\end{figure*}

In the left panel of Fig.~\ref{fig:comparison2} we report the same map
of $T_{\rm ew}$ shown on the left panel of Fig.~\ref{fig:com}, but
re-binned as the map of Fig.~\ref{fig:comparison}.

%
%
 
To highlight the temperature differences between the spectroscopic and
the emission-weighted temperature maps, in the right panel of
Fig.~\ref{fig:comparison2} we show the percentile difference of
$(T_{\rm ew}-T_{\rm spec})/T_{\rm spec}$. For better visualization we
only show the pixels where this difference is significant to at least
$3 \sigma$ confidence level, i.e. $|(T_{\rm ew}-T_{\rm
spec})/\sigma_{\rm spec}|>3 $, being $\sigma_{\rm spec}$ the 68 per
cent confidence level error associated with the spectroscopic
temperature measurement.  This plot clearly shows that there are many
regions where the difference between $T_{\rm ew}$ and $T_{\rm spec}$
is significant to better than $\ga 3\sigma$ confidence level.
Furthermore, for these pixels the discrepancy ranges from 50 per cent
to 200 per cent and even more. Of particular relevance are two cluster
regions showing a shock front in the emission-weighted map, in the
left panel of Fig.~\ref{fig:com}.  The map of the percentile
difference between \tew ~ and \tspec ~ shows discrepancies of 100-200
per cent, indicating that shock fronts predicted in the emission
weighted map are no longer detected in the observed spectroscopic map.

\begin{figure*} 
{\centering \leavevmode 
\psfig{file=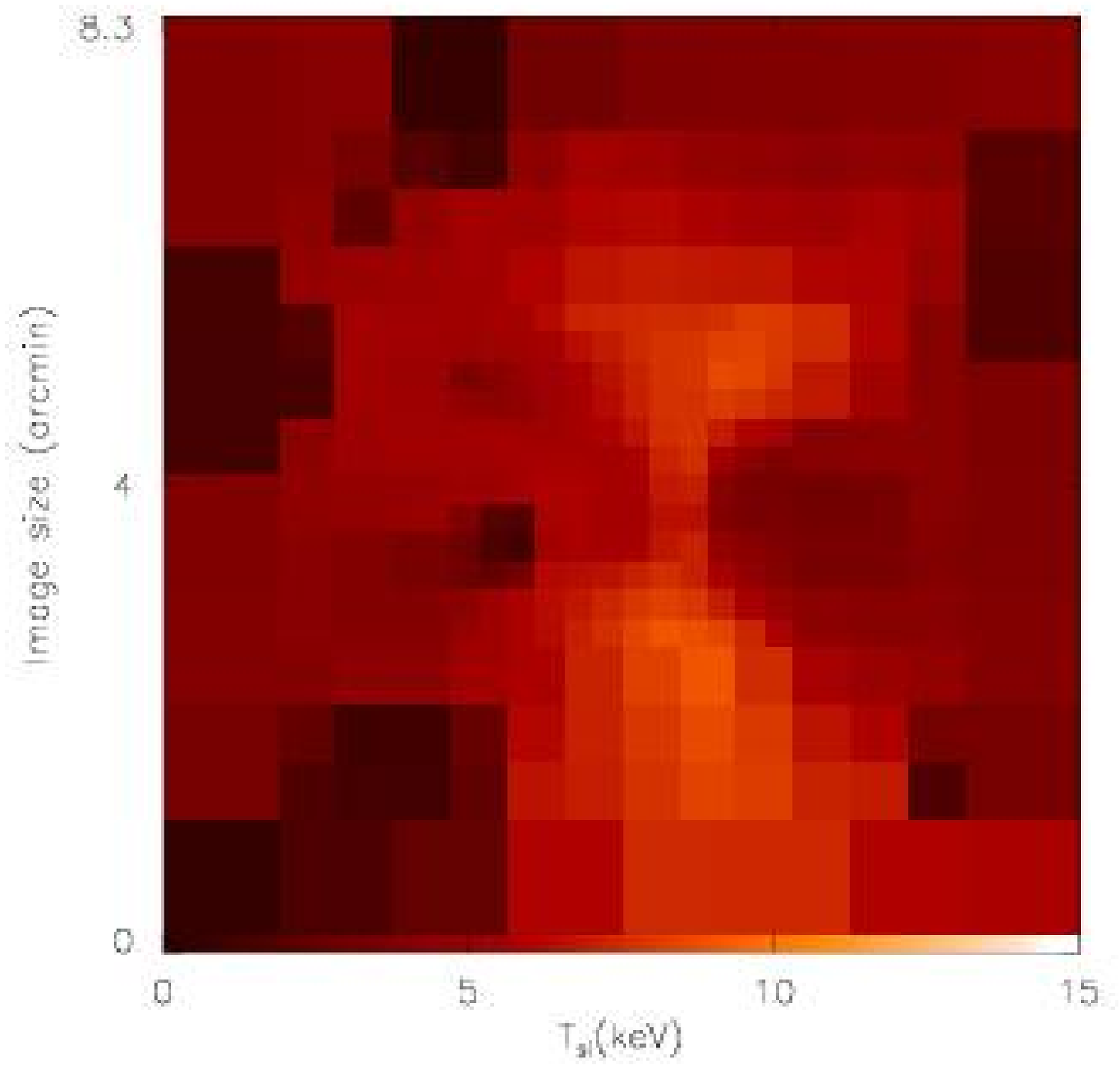,width=.45\textwidth} \hfil 
\psfig{file=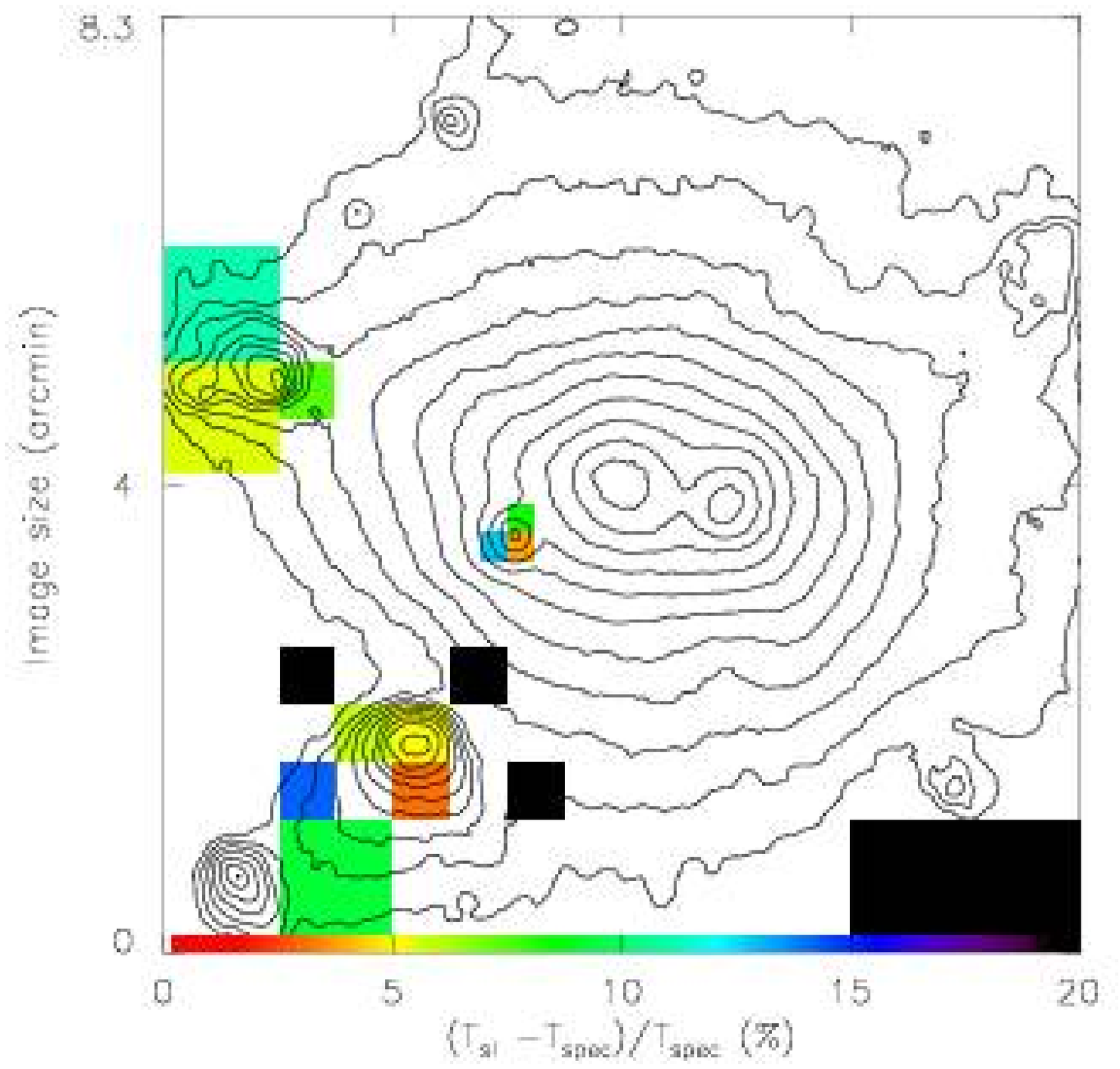,width=.45\textwidth} \hfil 
} 
\caption{Left panel: 
spectroscopic-like temperature map for the simulated cluster of
galaxies shown in the right panel of Fig.~\ref{fig:com} re-binned to
match the spatial resolution of the spectroscopic temperature map
shown in Fig.~\ref{fig:comparison}.  Right panel: percentile
difference between the spectroscopic and the spectroscopic-like
temperature maps.  In this map we show only the regions where the
significance level of the temperature discrepancy is at least $3
\sigma$, i.e.  $|(T_{\rm sl}-T_{\rm spec}) /\sigma_{\rm spec}|>3$,
where $\sigma_{\rm spec}$ is the 68 per cent confidence level error
associated to $T_{\rm spec}$.  }
\label{fig:comparison3} 
\end{figure*}

Conversely, in the left panel of Fig.~\ref{fig:comparison3} we show
the same map of $T_{\rm sl}$ shown in the right panel of
Fig.~\ref{fig:com}, but re-binned as the spectroscopic temperature map of
Fig.~\ref{fig:comparison}.  As we did for the emission-weighted
temperature map, in the right panel of Fig.~\ref{fig:comparison3} we
show the percentile difference of $(T_{\rm sl}-T_{\rm spec})/T_{\rm
spec}$.  Again we only visualize the pixels for which the difference
is significant to at least $3 \sigma$ confidence level.  The presence
in this map of fewer pixels clearly indicates that the match between
\tsl ~and \tspec ~ is much better than the one between \tspec ~ 
and \tew. Furthermore, most of these pixels shows very small
temperature discrepancies, being smaller than $10$ per cent. Only in 6
pixels we find a slightly higher temperature discrepancy, but in any
case smaller than $20$ per cent.
  
%
%
 
%
%
 
This demonstrates that the spectroscopic-like temperature gives a much
better estimate of the observed spectroscopic temperature than the
widely used emission-weighted one.  It is worth noting that in the
previous map 3 out of 6 pixels where the discrepancy is between 10 and
20 per cent correspond to cluster regions with very low surface
brightness.  We believe that the observed discrepancy in this case is
simply related to the very poor statistics used to determine
\tspec . In the other 3 cases the relatively higher discrepancy is
instead related to the fact that the thermal components of these
regions that contribute to the spectra have a quite large spread of
temperatures and, because the lower (dominant) thermal component is at
$T<3$~keV, \tspec ~ cannot be unequivocally identified: a proper
spectral analysis would require a fit with a two-temperature model.
  
It is very important to say that, unlike the emission-weighted, the
map on the right panel of Fig.~\ref{fig:comparison3} does not show big
discrepancies between \tsl ~ and \tspec ~ in both shock cluster
regions. This clearly indicate that \tsl ~ does a much better job than
\tew ~ in predicting the projected spectral properties of such peculiar 
thermal features.

\subsection {Cluster projected temperature profile} 
 
\begin{figure} 
{\centering \leavevmode
\psfig{file=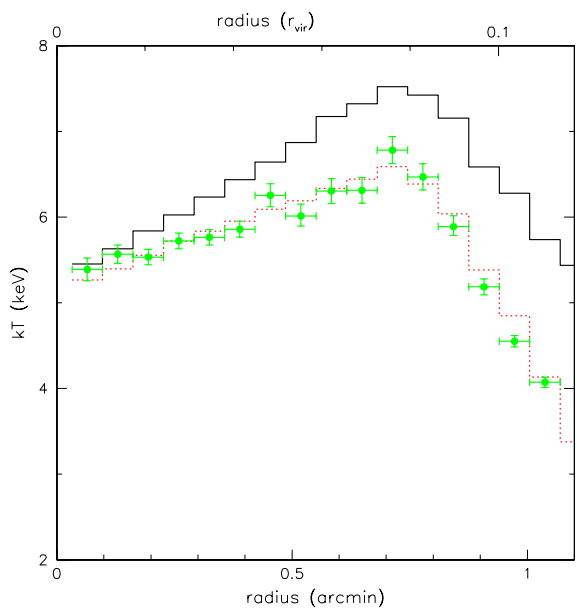,width=.49\textwidth} \hfil   
}  
\caption{ 
Temperature profiles of the simulated cluster of galaxies. Filled
circles refer to the mean spectroscopic temperature profile from
Gardini et al. (2004) and correspond to the region shown as a circle
in Fig.~\ref{fig:com}; error bars are at $68$ per cent confidence
level for one interesting parameter.  Solid and dotted histograms
refer to the mean emission-weighted and spectroscopic-like temperature
profiles directly extracted from the simulation.  }
\label{fig:profiles} 
\end{figure}

To conclude this section we also tested the accuracy of \tsl ~ in
predicting the observed spectroscopic temperature profile.  We again
used the spectroscopic projected temperature profile of
\citet{2003astro.ph.10844G}.  This was obtained by extracting spectra
from circular annuli centred on the cluster centre, out to the radius
identified by the circle in Fig.~\ref{fig:com}.  The size of the bin
was chosen in order to have approximately the same number of photons
inside each annulus.  The spectroscopic temperature profiles $T_{\rm
spec}$, together with their relative 68 per cent confidence level
errors $\sigma_{\rm spec}$, are shown as filled circles in
Fig.~\ref{fig:profiles}. In the same figure we show the
emission-weighted temperature and the spectroscopic-like temperature.
As already discussed in \citet{2003astro.ph.10844G}, we notice that
the emission-weighted temperature $T_{\rm ew}$ profile does not
reproduce the spectroscopic temperature profiles $T_{\rm spec}$. In
particular we confirm that the spectral temperatures are
systematically {\em lower} than the emission-weighted ones.
Conversely, the spectroscopic-like temperature profile provides a much
more accurate estimate, and falls within the error bars of the
``observed'' spectroscopic temperature profile everywhere but in two
annuli.

%
%
 
\section{Discussion and Conclusions}\label{par:dis}

In this paper we have studied the problem of performing a proper
comparison between temperatures obtained from the data analysis of
X-ray observations (i.e. projected spectroscopic temperatures, \tspec)
and temperatures derived directly from hydro-N-body simulations
(projected emission-weighted temperatures \tew).  In
\S~\ref{par:spec_prop} we show analytically that \tspec ~ is not a
well defined quantity.  In fact, it results from the fit of a
single-temperature thermal model to a multi-temperature source
spectrum; however, since the former cannot accurately reproduce the
spectral properties of the latter, it follows that \tspec ~ cannot be
unequivocally identified.  Generally speaking, this means that a
reliable comparison between simulations and observations can only be
done through the actual simulation of the spectral properties of the
simulated clusters: these properties will then be directly compared
with the observed ones.  Nevertheless, observed spectra are affected
by a number of factors that distort and confuse some of their
properties.  In some circumstances multi-temperature thermal source
spectra may appear statistically indistinguishable from a
single-temperature model.  In \S~\ref{par:spec_proj} we study this
aspect focusing our attention on observations of clusters of galaxies
made using the CCD detectors of \chandra ~ and \xmm . These detectors
are characterized by having both a similar intermediate energy
resolution and a similar energy response.  From our study we find two
very important results that for convenience we summarize below.

\begin{enumerate}
\item  Given a multi-temperature source spectrum, if the lowest dominant
temperature component has $T_1> 2-3$~keV, then a fit made with a 
single-temperature thermal model is statistically acceptable
regardless of the actual spread in temperature distribution. 

On the other hand, multi-temperature sources with $T_1<2$~keV will
most likely require multi-temperature spectral models.  This is
equivalent to say that \tspec ~ can be properly defined only for
spectra with $T_1>2-3$~keV. For lower temperatures the identification
of multi-temperature observed spectra with a single temperature is not
appropriate.  It is important to say that this result is intrinsically
related to the characteristics of the X-ray detector used for the
observation (e.g. its energy pass band and resolution) and not to a
possible inadequate photon statistics.  In fact our result was
obtained in the limit of high spectral photon number (see
\S~\ref{par:spec_proj}).  It is self-evident that this conclusion will
be true also for all the detectors whose spectral properties are
similar or worse than those on board of \chandra ~ and \xmm, while this
will not be true for X-ray spectrographs with either a much larger
energy pass band or a much higher energy resolution.

\item The emission-weighted temperature \tew, originally introduced to
provide a better comparison between simulations and observations, in
practice does not properly estimate the spectroscopic temperature
\tspec. In particular \tew ~ tends to overestimate \tspec . This
mismatch depends on the thermal inhomogeneity of the observed
multi-temperature source: the larger the spread in temperature of the
dominant components in the observed spectrum, the larger the
discrepancy.  In \S~\ref{par:slike} we derive a new formula, the
spectroscopic-like temperature function \tsl ~(see
Eq.~\ref{eq:new_form_TW}), and show that \tsl~  can approximate \tspec ~
to a level better than 10 per cent regardless of the temperature
spread.  It is worth noticing that  \tsl ~ weights each thermal
component directly by the emission measure but inversely by their
temperature to the power of $3/4$.  This explains why the observed
\tspec ~ is biased toward the lower values of the dominant thermal
components.  

\end{enumerate}

%
%

\begin{figure}
{\centering \leavevmode
\psfig{file=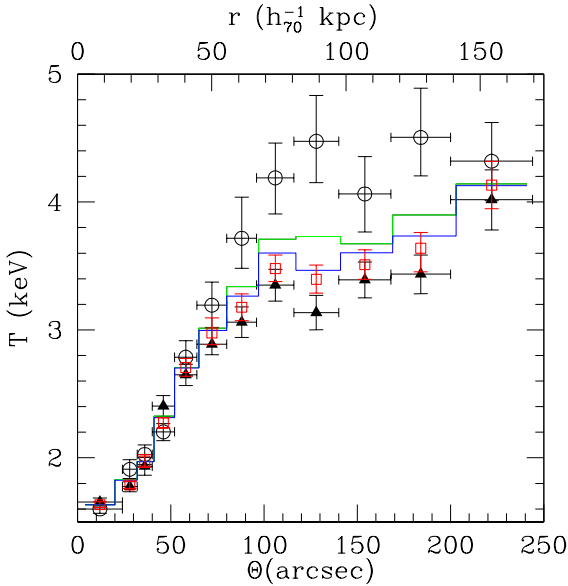,width=.45\textwidth}
}
\caption{Temperature profiles of the galaxy cluster
2A~0335 (from \citealt{2003ApJ...596..190M}). Filled triangles and
open circles refer to the projected cluster temperature profile in the
northern and southern sectors, respectively. Open squares indicate the
overall projected cluster temperature profile obtained from a circular
radial analysis.  The upper and the lower histograms show the profiles
for the emission-weighted and spectroscopic-like temperatures,
respectively.  These histograms have been obtained by combining the
northern and southern temperature profiles.}
\label{fig:2A0335}
\end{figure}

These two results have important observational and theoretical
implications for the study of X-ray clusters of galaxies.  First of all
the equivalence between observed ``hot'' multi-temperature spectra and
single-temperature models implies that, using the CCD detectors of
\chandra ~ and \xmm , it is observationally impossible to disentangle
the single or multi-temperature nature of any observed
spectrum whose projected temperature is higher than $3$~keV by
performing a simple overall X-ray spectral analysis.  In addition, the
temperature derived from this spectral analysis is not the
emission-weighted value, but it is biased toward the lowest dominant
thermal component of the overall spectrum.  The consequences for
studies of temperature profiles in clusters are immediate. In
virtually all works on the subject, in fact, the cluster temperature
profile is derived by extracting spectra from concentric circular or
elliptical annuli centred on the cluster X-ray peak. Together with the
cluster gas distribution, this temperature profile is used to estimate
the cluster mass by assuming hydrostatic equilibrium. It is
self-evident that, if the cluster gas temperature distribution is
azimuthally asymmetric, the temperature profile derived from the
radial analysis is biased toward lower temperature values and so does
the estimated mass.  To better show this aspect we discuss the results
of the data analysis of the cluster of galaxies \2A ~ selected, as
example, among the many published in the literature.  In particular, all
the temperature measurements discussed below are taken from
\cite{2003ApJ...596..190M}.  In Fig.~\ref{fig:2A0335} we compare the
projected temperature profiles of the cluster of galaxies \2A ~
extracted from different sectors.  Filled triangles and open circles
indicate the temperature profiles obtained from the Northern (from
-90\degd ~ to 90\degd ; angles are measured from North toward East)
and Southern (from 90\degd ~ to 270\degd ) sectors of the clusters.
As highlighted by \cite{2003ApJ...596..190M}, the temperature profile
of \2A ~ is clearly azimuthally asymmetric. We notice that in the
100-200 arcsec radial interval the temperature profile of the southern
sector is between 30 and 50 per cent hotter than the Northern one.
The open squares in Fig.~\ref{fig:2A0335} indicate the cluster
temperature profile obtained from the circular radial analysis.  For
convenience we report as histogram the expected temperature profile
obtained by combining the Northern and Southern temperature profiles
using the emission-weighted formula.  From this figure it is evident
that, consistently with what discussed so far, the circular radial
temperature profile is significantly lower that the expected
emission-weighted profile and is biased toward the values of the
Northern profile.  For completeness in the same figure we added as
histogram the temperature profile obtained by combining the Northern
and Southern temperature measurements using the spectroscopic-like
formula.  This latter profile is perfectly consistent with the
measured one proving once more that, unlike the emission-weighted
temperature, our spectroscopic-like formula provides a very good
approximation to the spectroscopic temperature measurements.

It is important to say that, by construction, the difference between
\tew ~ and \tsl ~ mainly depends on the complexity of the projected
thermal structure of the cluster gas.  It is clear that the findings
of this study are especially relevant for major-merger clusters rather
than for relaxed clusters. They may have also important implications
for the study of all structures with very strong temperature gradients
like, for example, the shock fronts.  In fact, because of the
temperature bias of \tspec , shock fronts in real observations appear
much weaker than the predictions of virtually all the
emission-weighted temperature map published in literature.  This has
been shown in \S~\ref{par:temap} and is evident in
Fig.~\ref{fig:com}. From this figure we immediately see that the two
shock fronts clearly visible in the emission-weighted map are no
longer detected in the spectroscopic-like temperature map.  This
temperature bias may explain why, although simulations predict that
shock fronts are quite common in clusters of galaxies, to date we have
very few observations of clusters in which they are clearly present.

To conclude we stress once more that the emission-weighted temperature
function may give a misleading view of the actual gas temperature
structure as obtained from X-ray observations. Thus, since the
emission-weighted temperature has no physical meaning (unlike the
mass-weighted one), here we propose to theoreticians and N-body
simulators to finally discard its use.  We remind that great attention
must be paid when comparing simulations and X-ray observations.  Under
the most generic conditions such comparison can only be done through
the actual simulation of the spectral properties of the simulated
clusters, so that software packages like X-MAS
\citep{2003astro.ph.10844G} become fundamental.  Nevertheless, if the
cluster temperature is sufficiently high and if the spectral
properties of the detector used for the observation are similar or
worse that the ones of \chandra ~ and \xmm ~ we showed that our
proposed spectroscopic-like temperature function may be considered 
an appropriate tool for the job.

\section*{Acknowledgments} 
PM, LM, and GT are grateful to the Aspen Center for Physics, where the
idea for this paper comes out.  We thank Stefano Borgani and Klaus
Dolag for clarifying discussions.  This work was partially supported
by European contract MERG-CT-2004-510143 and CXC grants GO2-3177X.

\end{document}